\documentclass[]{pasj01}
\usepackage{color}

\begin{document} 
\Received{}
\Accepted{}

\title{On the different levels of dust attenuation to nebular and stellar light in star-forming galaxies}


\author{Yusei \textsc{Koyama}\altaffilmark{1,2}}%
\altaffiltext{1}{Subaru Telescope, National Astronomical Observatory of Japan, National Institutes of Natural Sciences, 650 North A'ohoku Place, Hilo, HI 96720, U.S.A.}
\altaffiltext{2}{Graduate University for Advanced Studies (SOKENDAI), Osawa 2-21-1, Mitaka, Tokyo 181-8588, Japan}
\email{koyama@naoj.org}

\author{Rhythm \textsc{Shimakawa}\altaffilmark{1}}

\author{Issei \textsc{Yamamura}\altaffilmark{3}}
\altaffiltext{3}{Institute of Space Astronautical Science, Japan Aerospace Exploration Agency, Sagamihara, Kanagawa 252-5210, Japan}

\author{Tadayuki \textsc{Kodama}\altaffilmark{4}}
\altaffiltext{4}{Astronomical Institute, Tohoku University, 63 Aramaki, Aoba-ku, Sendai 980-8578, Japan}

\author{Masao \textsc{Hayashi}\altaffilmark{5}}
\altaffiltext{5}{National Astronomical Observatory of Japan, Osawa 2-21-1, Mitaka, Tokyo 181-8588, Japan}



\maketitle

\begin{abstract}
As a science verification study of the newly released AKARI/FIS Faint Source Catalog ver.1, this paper discusses the different levels of dust attenuation toward stellar light and nebular emission lines within local star-forming galaxies at 0.02$<$$z$$<$0.10. By constructing an updated version of the AKARI--SDSS--GALEX matched galaxy catalog (with $>$2,000 sources), we compare the dust attenuation levels toward stellar light (from $L_{\rm IR}$/$L_{\rm UV}$ ratio) and nebular emission lines (from H$\alpha$/H$\beta$ ratio). We find that there is a clear trend that more massive galaxies tend to have higher ``extra'' attenuation toward nebular regions, while galaxies with higher specific star formation rates tend to have lower extra attenuation. We also confirm these trends by using the WISE mid-infrared photometry with a significantly large sample size of the WISE--SDSS--GALEX galaxies ($>$50,000 sources). Finally, we study how the levels of extra attenuation toward nebular regions change across the SFR--$M_{\star}$ plane. We find that, even at a fixed stellar mass, galaxies located {\it below} the main sequence tend to have higher levels of extra attenuation toward nebular regions, suggesting the change in dust geometry within the galaxies across the star-forming main sequence during the course of star formation quenching process.

\end{abstract}


\section{Introduction}

Measuring the star formation rate (SFR) of galaxies is always an important task in the extra-galactic astronomy. A common approach to derive SFR is to observe the rest-frame UV light from young massive stars, or to measure the nebular emission line fluxes such as H$\alpha$ ($\lambda$$=$6563\AA) line associated with young star-forming regions, applying a proper dust attenuation correction (e.g.\ \cite{Kennicutt1998}; \cite{Kennicutt2009}; \cite{Hao2011}; \cite{Kennicutt2012}). Because star-forming regions are dust-rich environment in general, the light emitted from those young star-forming (H{\sc ii}) regions are known to be heavily obscured by dust, and unfortunately, the dust attenuation correction has been one of the biggest sources of uncertainty when deriving the SFRs.

The H$\alpha$ line is recognized as one of the best indicators of SFR. The dust attenuation at H$\alpha$ line can ideally be estimated by observing H$\beta$ ($\lambda$$=$4861\AA) line at the same time and translating the H$\alpha$/H$\beta$ line flux ratio (Balmer decrement) to the dust attenuation levels with an assumption of reddening curves. Although it is true that even H$\alpha$ line could be heavily obscured in the case of extremely dusty sources (e.g.\ \cite{Poggianti2000}; \cite{Koyama2010}), it is widely recognized that the dust-corrected H$\alpha$-based SFR is one of the most reliable estimators of SFR. The H$\alpha$ line has been commonly used for SFR measurements of local galaxies, and with the recent improvements of near-infrared observational technologies, it has now become much easier to access to the (redshifted) H$\alpha$ line of distant galaxies across environment (e.g.\ \cite{Geach2008}; \cite{Garn2010a}; \cite{Sobral2013}; \cite{Koyama2013}; and many others). 

However, because H$\beta$ line is much fainter than H$\alpha$, it is still very challenging to directly measure the dust attenuation to H$\alpha$ line for a statistical sample of galaxies at high redshifts with the Balmer decrement; indeed, only a small fraction of high-redshift works successfully detect both H$\alpha$ and H$\beta$ lines of individual sources for a statistical sample of galaxies (e.g.\ \cite{Shivaei2015b}; \cite{Reddy2015}). An alternative approach exploited in many studies is to use the dust reddening derived from the broad-band SED fitting to predict the dust attenuation to H$\alpha$ line. However, in addition to the well-known degeneracy between age, metallicity, and dust reddening in the broad-band SEDs, this process accompanies a large uncertainty, because the dust reddening estimated from broad-band SEDs indicates the reddening toward {\it stellar} light, while the reddening toward {\it nebular} emission lines could be different (e.g.\ \cite{Reddy2015}; \cite{Koyama2015}; \cite{Zahid2017}).

It should be noted that nebular emission lines from H{\sc ii} regions tend to be more obscured by dust than the stellar continuum light averaged over the galaxies. \citet{Calzetti1997} showed that stellar continuum light are a factor of $\sim$2$\times$ {\it less} reddened than the nebular emission lines on average, and established a relation between color excess toward stellar continuum and nebular emission lines with an equation of E(B$-$V)$_{\rm star}$$=$0.44$\times$E(B$-$V)$_{\rm gas}$ (see also \cite{Calzetti1994}; \cite{Calzetti2000}). Although many studies apply this relation to convert the reddening to stellar light to that of ionized gas until recently, it should be noted that the factor ``0.44'' in the above equation is derived as an average value of a wide variety of galaxy population in the local universe (from starbursts to dwarfs; see \cite{Calzetti1997}), and it would not be a number applicable to all types of galaxies.  

Some recent studies suggest that the factor of this ``selective'' attenuation (or ``extra'' attenuation) toward nebular emission lines is different for distant galaxies (e.g. \cite{Erb2006}; \cite{Reddy2010}; \cite{Kashino2013}; \cite{Pannella2015}), and a growing number of high-$z$ studies do not blindly assume the same relation as for local galaxies (e.g.\ \cite{Forster-Schreiber2009}; \cite{Wuyts2013}; \cite{Tadaki2013}; \cite{Theios2018}). It is often argued that the different levels of attenuation toward ionized gas is originated by the fact that stars and ionized gas do not occupy the same regions within the galaxies (e.g.\ \cite{Calzetti1994}; \cite{Wild2011}; \cite{Price2014}; \cite{Reddy2015}). The ionized gas requires the presence of young massive (short-lived) stars to ionize themselves (hence they are always located in the birth clouds), while stellar continuum light can also be contributed by long-lived, non-ionizing stars distributed throughout the galaxy; so-called ``two-component'' dust model composed of birth-clouds and diffuse ISM as described by \citet{Charlot2000}. It is of great interest to understand what determines the levels of extra attenuation toward nebular regions---this is not only to reduce the uncertainties of SFR estimates, but also to understand the link between star/gas geometry inside the galaxies along the process of galaxy evolution.

\begin{figure*}
 \begin{center}
\includegraphics[width=7.8cm,angle=0]{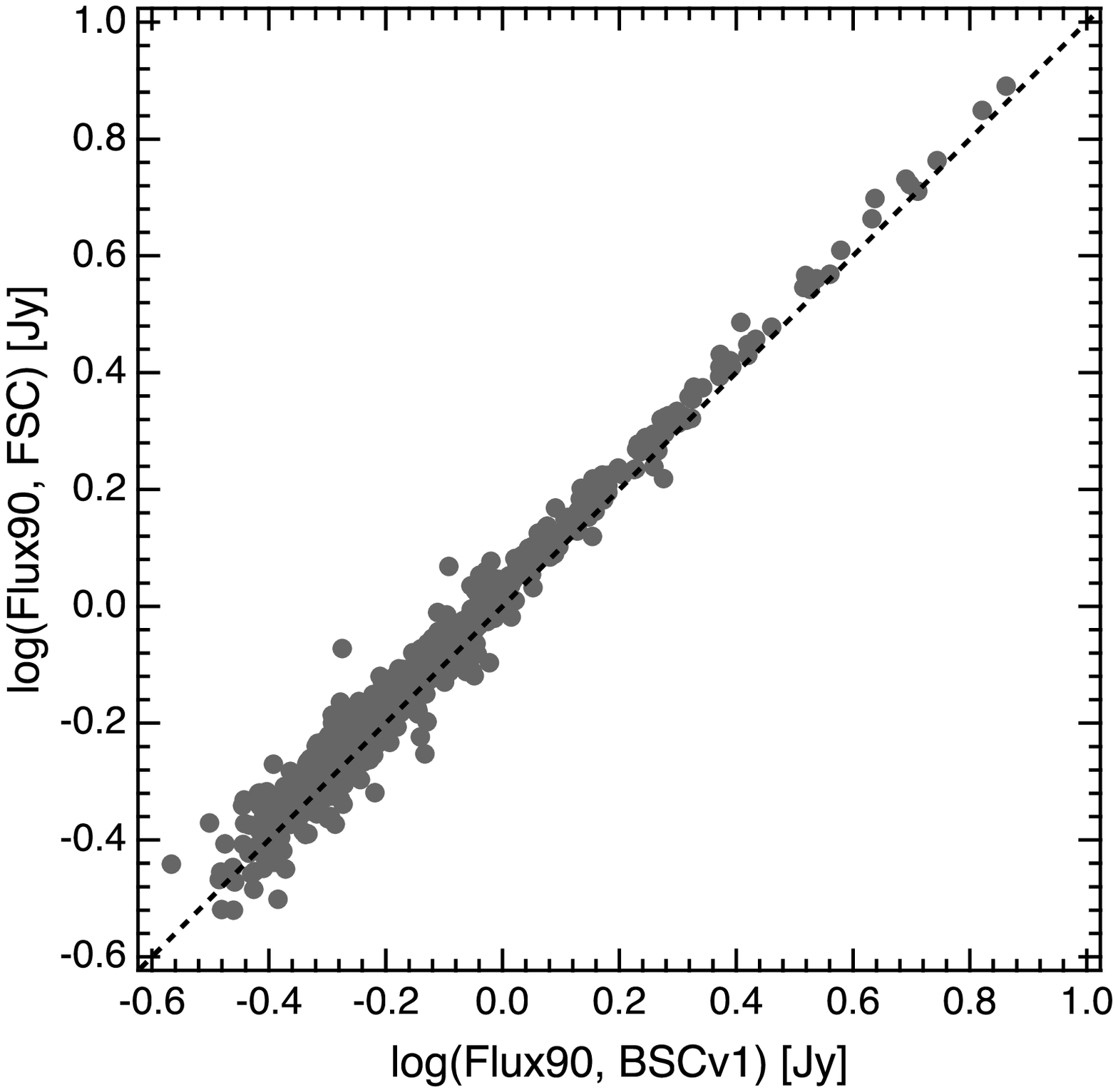} 
\includegraphics[width=7.8cm,angle=0]{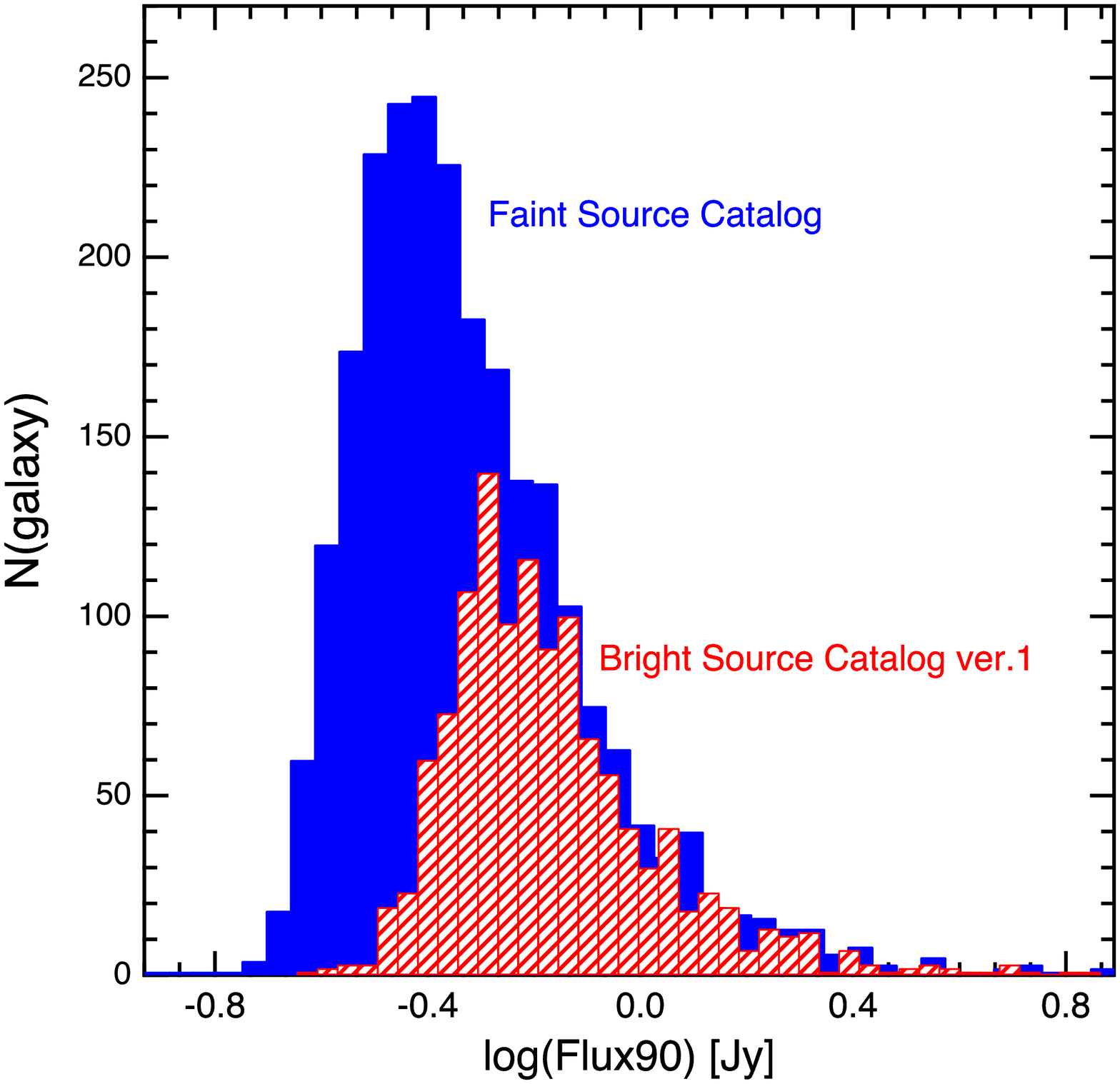} 
 \end{center}
\vspace{-3mm}
\caption{(Left): A comparison between AKARI/FIS WIDE-S (90~$\mu$m) flux densities from the Bright Source Catalog ver.1 (BSCv1) and Faint Source Catalog ver.1 (FSCv1) for star-forming galaxies at 0.02$<$$z$$<$0.1 identified in both of the catalogs, showing a reasonable agreement between BSCv1 and FSCv1 over a wide flux range. (Right): The distribution of 90~$\mu$m flux densities (Flux90) for our old/new AKARI--SDSS--GALEX star-forming galaxy samples. The red (hatched) histogram shows the distribution of Flux90 for the AKARI(BSCv1) sample presented in our previous work (\cite{Koyama2015}), while the blue (filled) histogram shows the result for our new AKARI(FSCv1) sample. This plot demonstrates that the number of our AKARI--SDSS--GALEX galaxy sample has been significantly increased mainly at the faint end with the improved sensitivity of AKARI Faint Source Catalog. }\label{fig:BSC_FSC_compare}
\end{figure*}

A method to study the extra attenuation toward nebular regions would be to compare the E(B$-$V)$_{\rm gas}$ and E(B$-$V)$_{\rm star}$ measured independently. The E(B$-$V)$_{\rm gas}$ can be measured with H$\alpha$/H$\beta$ ratio as described above, and the E(B$-$V)$_{\rm star}$ can be estimated from UV attenuation measured by far-infrared (FIR) to UV luminosity ratio with an assumption of attenuation curve. 
The key observational data is therefore deep MIR--FIR photometry covering a wide area on the sky to directly trace the dust thermal emission from galaxies. The Wide-field Infrared Survey Explorer ({\it WISE}; \cite{Wright2010}) performed all-sky survey at the MIR (3.4, 4.6, 12, 22~$\mu$m; W1--W4), and provides an excellent tool for this purpose. Also, the {\it AKARI} (\cite{Murakami2007}) Far-Infrared Surveyor (FIS; \cite{Kawada2007}) performed all-sky survey at 65, 90, 140, and 160~$\mu$m with high sensitivity and high angular resolution (\cite{Doi2015}; \cite{Takita2015}), and thus serves as another ideal tool for this purpose. 

In \citet{Koyama2015}, we constructed an {\it AKARI--SDSS--GALEX} star-forming galaxy sample, by cross-matching the sources in the AKARI-FIS Bright Source Catalog ver.1 (\cite{Yamamura2010}) to SDSS (DR7) and GALEX (GR5) data. By comparing the $A_{\rm UV}$ (derived from $L_{\rm IR}$/$L_{\rm UV}$ ratio) and $A_{\rm H\alpha}$ (derived from Balmer decrement), we find a hint that more massive galaxies and/or galaxies with low H$\alpha$ equivalent widths (EW$_{\rm H\alpha}$) tend to show higher levels of extra attenuation toward nebular regions. In this paper, as a science demonstration study of the newly released AKARI Faint Source Catalog ver.1 (\cite{Yamamura2018}), we will revisit the issue of this extra attenuation toward nebular regions by constructing an updated version of the {\it AKARI--SDSS--GALEX} matched galaxy catalog. 

This paper is organized as follows. In Section~2, we present our new {\it AKARI--SDSS--GALEX} star-forming galaxy catalog with the newly released AKARI Faint Source Catalog (Section~2.1), and derive their FUV-, H$\alpha$-, and FIR-based SFRs as well as the dust attenuation levels toward stellar and nebular emission (Section~2.2--2.4). In Section~3, we investigate how the ``extra'' attenuation toward nebular regions changes with various galaxy properties using the {\it AKARI--SDSS--GALEX} sample. In Section~4, we verify our results by using {\it WISE} data as it also covers all sky and provides an independent measurement of IR-based SFRs. In Section~5, by using our {\it AKARI--SDSS--GALEX} and {\it WISE--SDSS--GALEX} sample, we show how the levels of extra attenuation to nebular regions changes across the star-forming main sequence. In Section~6, we describe caveats on the interpretation of our results by comparing the results when we use H$\alpha$-based SFRs instead of UV/IR-based SFRs. Finally, our conclusions are summarized in Section~7. Throughout the paper, we adopt the cosmological parameters of $\Omega_{\rm{M}} =0.3$, $\Omega_{\Lambda} =0.7$, and $H_0 =70$ km s$^{-1}$Mpc$^{-1}$, and we assume the \citet{Kroupa2001} initial mass function (IMF). All magnitudes are given in the AB system.

\section{New AKARI-SDSS-GALEX sample}
\label{sec:data}

\subsection{Sample selection}

In \citet{Koyama2015}, we constructed a star-forming galaxy sample ($N$$=$1,200) in the local universe by matching the {\it SDSS} Data Release 7 (DR7; \cite{Abazajian2009}) spectroscopic galaxy sample, UV sources from the All-sky Imaging Survey of {\it GALEX} fifth data release (GR5), and FIR sources from the AKARI FIS Bright Source Catalog ver.1 (BSCv1; \cite{Yamamura2010}). We refer the readers to \citet{Koyama2015} for details of the source matching procedure, but we here briefly describe the process of our sample selection. We note that the dataset is the same as those used in \citet{Koyama2015}, except that we replace our FIR data from the AKARI FIS Bright Source Catalog ver.1 to the newly released AKARI Faint Source Catalog ver.1 (FSCv1; \cite{Yamamura2018}). The key observational quantities used in this paper are SDSS H$\alpha$ and H$\beta$ fluxes to derive H$\alpha$-based SFRs (Section~2.2), and GALEX FUV and AKARI FIR (90$\mu$m and 140$\mu$m) photometry to derive UV$+$FIR SFRs (Section~2.3).

The first step is to match the SDSS (DR7) spectroscopic catalog with the FUV-detected sources selected from GALEX (GR5) All-sky Imaging Survey (AIS) with a help of the SDSS (DR7)--GALEX (GR5) matched photometric catalog published by \citet{Bianchi2011}\footnote{https://archive.stsci.edu/prepds/bianchi\_gr5xdr7/}. We use the Max Planck Institute for Astrophysics and Johns Hopkins University (MPA/JHU) catalog\footnote{https://wwwmpa.mpa-garching.mpg.de/SDSS/DR7/} for spectroscopic measurements including the Galactic extinction correction, stellar absorption correction for emission line fluxes, and stellar mass estimates for individual galaxies with SED fitting (\cite{Kauffmann2003a}; \cite{Salim2007}). We restrict the sample to those within the redshift range of $0.02 < z < 0.10$, and we also apply a stellar mass cut of $\log (M_{\star}/M_{\odot}$)$>$8.5. We further require detection of four major emission lines (H$\alpha$, [NII], H$\beta$, [OIII]) to select star-forming galaxies using the 'BPT' diagram (\cite{Baldwin1981}), following the standard criteria to distinguish star-forming galaxies from AGNs (\cite{Kauffmann2003b}; \cite{Kewley2006}). We thus select 78,731 {\it SDSS--GALEX} sources satisfying all the above criteria.

We then match our {\it SDSS--GALEX} galaxy sample with the AKARI FIS Faint Source Catalog ver.1 (hereafter FSCv1; \cite{Yamamura2018}). The AKARI FSCv1 contains 401,157 point sources selected with a different detection policy from the AKARI FIS Bright Source Catalogs (ver.1/ver.2) to improve the sensitivity, particularly in the high ecliptic latitude regions where the number of scans is large (see \cite{Yamamura2018} for details). We note that the majority of the AKARI sources are Galactic objects distributed along the Galactic plane, and we find that $\sim$15,000 galaxies are located within the SDSS survey area. Considering the PSF size of {\it AKARI} at 90$\mu$m ($\sim$40-arcsec), we use 20-arcsec radius aperture to search counterparts, and find 2,416 star-forming galaxies identified in all the {\it AKARI}, {\it SDSS}, and {\it GALEX} catalogs, with a median redshift of $z$$=$0.043, and with a median separation of 5.1 arcsec (measured as angular distance between the positions of {\it AKARI} and {\it SDSS} coordinates). 

In Fig.~\ref{fig:BSC_FSC_compare} (left), we compare the AKARI WIDE-S (90$\mu$m) photometry from AKARI/FIS BSCv1 and FSCv1 for galaxies identified in both of the catalogs. More detailed comparison between the catalogs will be presented in \citet{Yamamura2018}, but we here confirm that the FIR flux measurements agree well between the BSCv1 and FSCv1 over a wide flux range for local star-forming galaxies. We also show in Fig.~\ref{fig:BSC_FSC_compare} (right) the 90$\mu$m flux distribution of the {\it AKARI--SDSS--GALEX} sample from our previous version (with BSCv1; red hatched histogram) and the updated version of the catalog (with FSCv1; blue filled histogram). It is clear that the increase of the total sample size mainly comes from the improved sensitivity at $\log F_{\rm 90\mu m} \lesssim -$0.4~[Jy].

\begin{figure}
\vspace{0mm}
 \begin{center}
\includegraphics[width=8.0cm,angle=0]{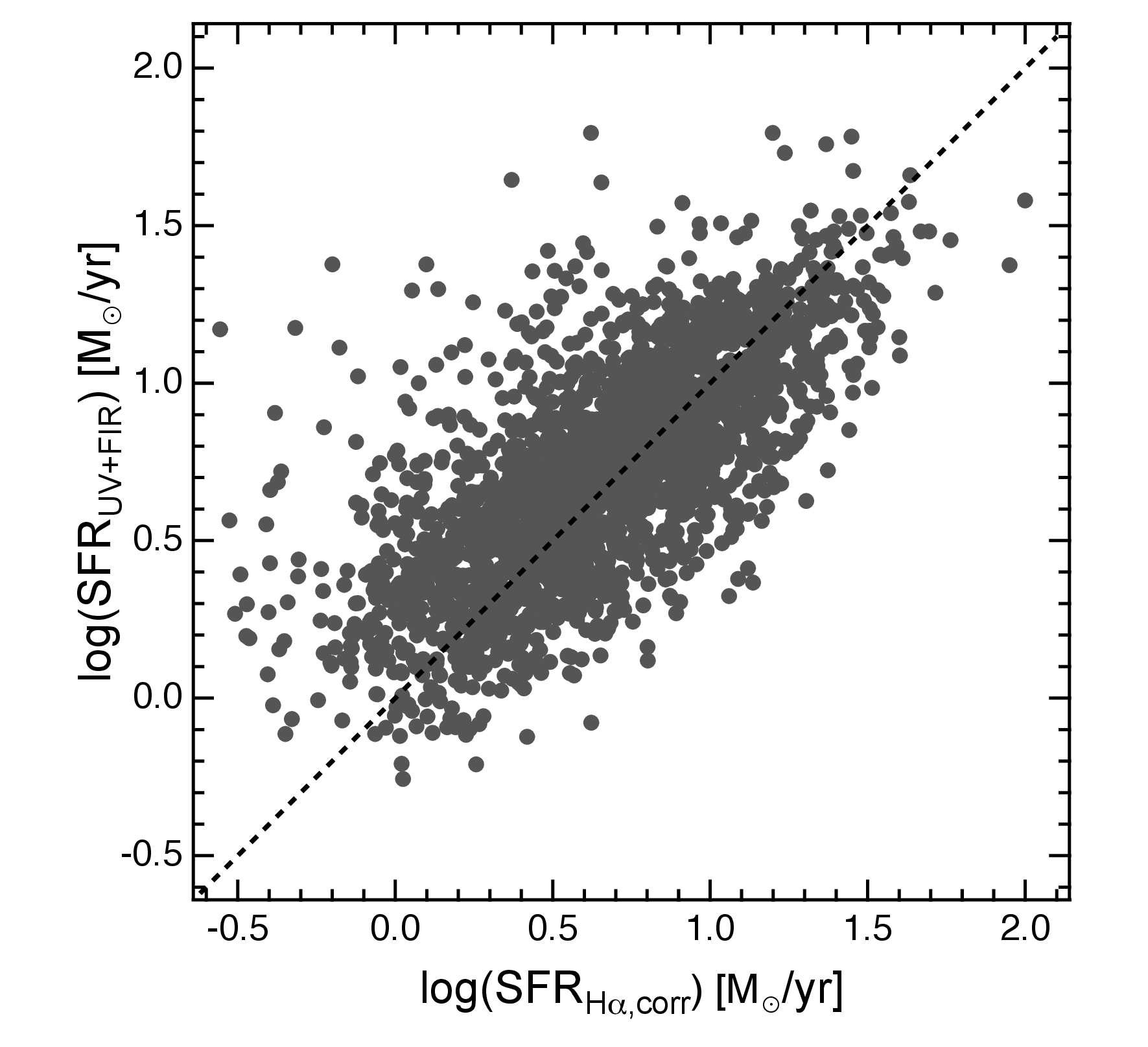} 
 \end{center}
\vspace{-3mm}
\caption{SFRs derived from UV and FIR photometry (SFR$_{\rm UV+FIR}$) of our new AKARI--SDSS--GALEX sample plotted against their H$\alpha$-based SFRs with the corrections for fiber/aperture effects as well as dust attenuation effect (SFR$_{\rm H\alpha, corr}$; see text for details). As shown by \citet{Koyama2015} with the older version of the catalog, this plot demonstrates good agreement between the two independently measured SFRs. }\label{fig:SFRHa_SFRFIR_compare}
\end{figure}
\begin{figure}
\vspace{-2mm}
 \begin{center}
\includegraphics[width=8.1cm,angle=0]{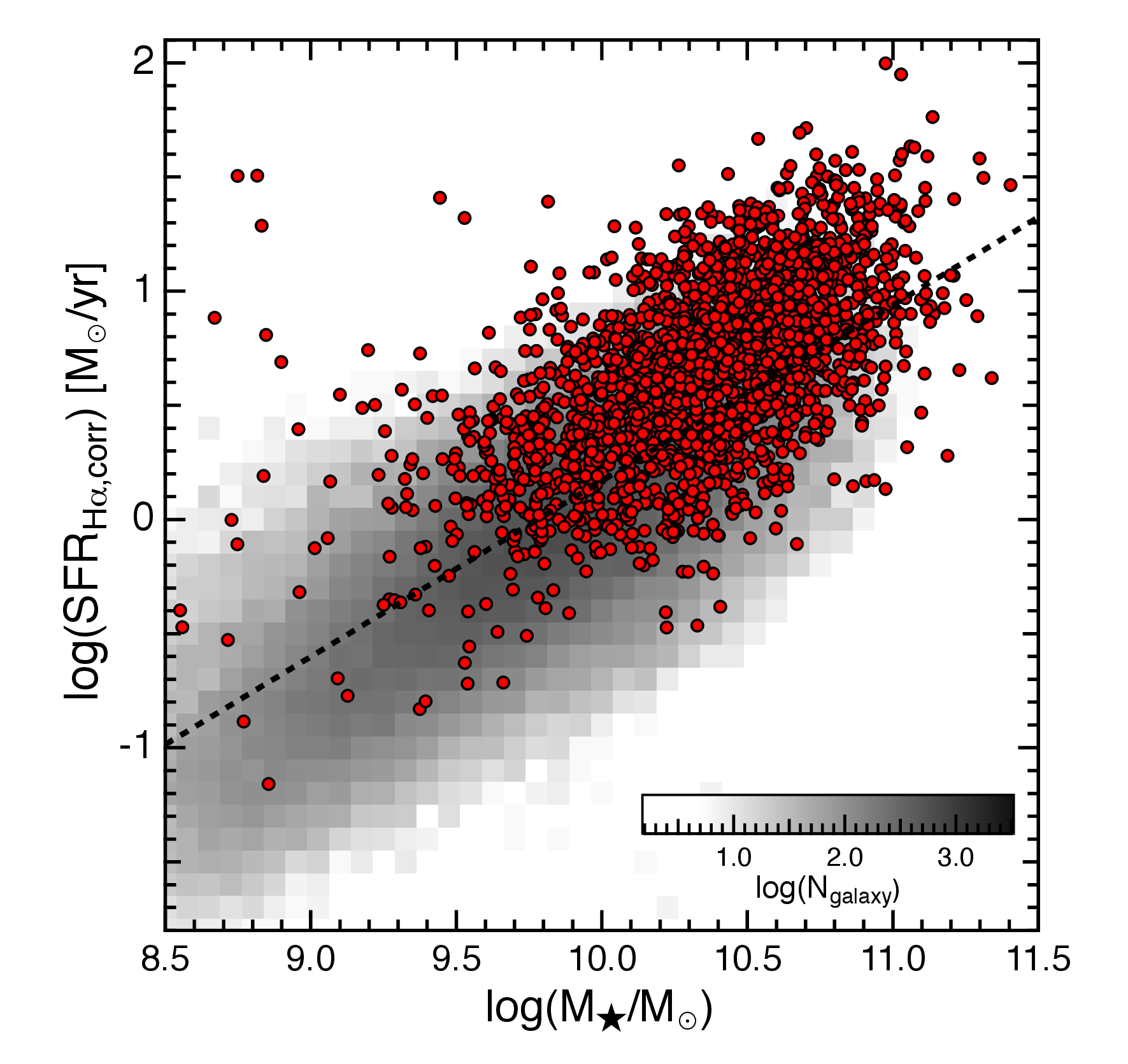} 
 \end{center}
\vspace{-3mm}
\caption{The AKARI(FSCv1)--SDSS--GALEX galaxy sample plotted on the SFR--$M_{\star}$ diagram (red circles). The gray-scale density plot indicates the distribution of all the SDSS--GALEX matched SF galaxy sample. To create this plot, we divide the panel into 40$\times$40 sub-grids, and compute the number counts of galaxies in each pixel. We note that the $M_{\star}$ and SFR$_{\rm H\alpha, corr}$ are derived in the same way for both AKARI-detected and undetected sources.}\label{fig:sample_on_SFMS}
\end{figure}
\begin{figure*}
\vspace{-4mm}
 \begin{center}
  \includegraphics[width=17.5cm,angle=0]{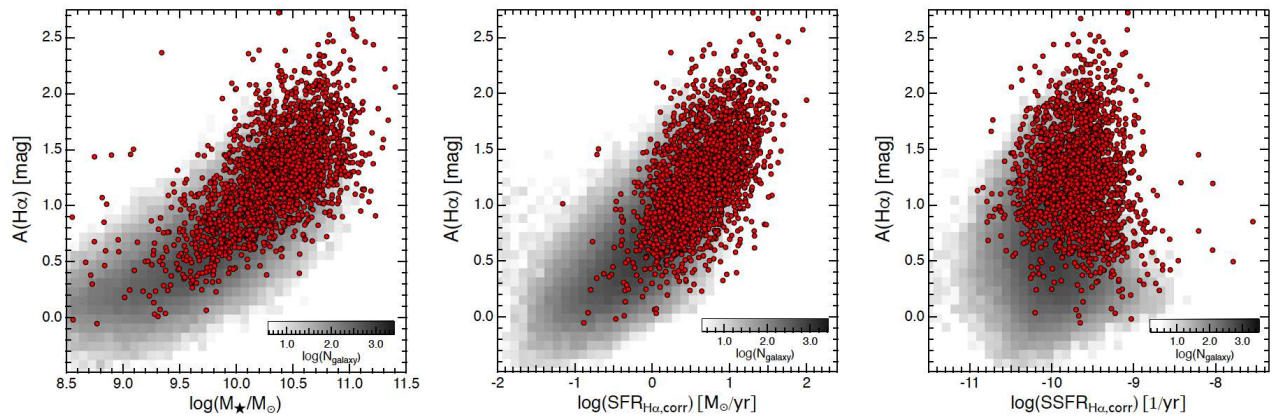} 
 \end{center}
\vspace{-3mm}
\caption{Dust attenuation at H$\alpha$ ($A_{\rm H\alpha}$) measured with H$\alpha$/H$\beta$ flux ratio (Balmer decrement) plotted against stellar mass (left), SFR (middle), and specific SFR (right). The red circles indicate our new AKARI--SDSS--GALEX sample, while the gray-scale density plots show the overall distribution of SDSS--GALEX matched SF galaxy sample. It can be seen that our AKARI-detected galaxies tend to be massive, highly star-forming population, and that they are heavily obscured by dust for their specific SFR.}\label{fig:AHa_vs_properties}
\end{figure*}

\subsection{H$\alpha$-based SFRs and dust attenuation toward nebular regions}

The key physical parameters that we use in this study are H$\alpha$- and UV/FIR-based SFRs and dust attenuation toward stellar and nebular light for individual galaxies. We derive these quantities basically following the same way as presented in \citet{Koyama2015}, but here we provide a quick overview of the procedure.  

To derive the H$\alpha$-based SFRs, we first calculate the dust attenuation at H$\alpha$ line by using the H$\alpha$/H$\beta$ flux ratio (Balmer decrement) with the following equation, assuming the Case B recombination at $T$$=$10$^4$~K with an electron density of $n_{\rm e}$$=$10$^2$ cm$^{-3}$:
\begin{equation}
A_{\rm H\alpha}=\frac{-2.5k_{\rm H\alpha}}{k_{\rm H\alpha}-k_{\rm H\beta}} \log \left( \frac{2.86}{F_{\rm H\alpha}/F_{\rm H\beta}} \right), 
\end{equation}
\begin{equation}
E(B-V)_{\rm gas} = A_{\rm H\alpha} / k_{\rm H\alpha}. 
\end{equation}
We here assume the \citet{Cardelli1989} Galactic extinction curve to determine the $k_{\rm H\alpha}$ and $k_{\rm H\beta}$ in the above equations, following the original derivation of the \citet{Calzetti2000} relation (see e.g.\ \cite{Steidel2014}; \cite{Reddy2015}; \cite{Shivaei2015a}). We emphasize that our conclusions are unaffected even if we apply the \citet{Calzetti2000} curve to both nebular and stellar light as assumed by many studies including our previous work (e.g. \cite{Garn2010b}; \cite{Sobral2012}; \cite{Koyama2015}), while it would bring a small systematic offset in the estimate of $A_{\rm H\alpha}$ by $\sim$10\%. 

We derive the H$\alpha$-based SFRs (SFR$_{\rm H\alpha}$) using the \citet{Kennicutt1998} calibration assuming \citet{Kroupa2001} IMF\footnote{We follow \citet{Kennicutt2009} to reduce the SFR by a factor of 1.44 to convert the SFRs derived under the assumption of \citet{Salpeter1955} IMF (i.e. the original \cite{Kennicutt1998} prescription) to that derived with \citet{Kroupa2001} IMF. We apply the same (a factor of 1.44) scaling when we derive SFR$_{\rm UV}$ and SFR$_{\rm FIR}$ in this work.}:
\begin{equation}
\rm{SFR_{\rm H\alpha}} = 5.5 \times 10^{-42} L_{\rm H\alpha} {\rm [erg\,s^{-1}]}. 
\end{equation}
We here correct for the dust attenuation effect based on the $A_{\rm H\alpha}$ estimated above. We also apply aperture correction to the H$\alpha$ flux based on the difference between the total (petrosian) magnitudes and SDSS fiber magnitudes at $r$-band (i.e.\ $m_{r, {\rm petro}}$$-$$m_{r, {\rm fiber}}$), assuming that the dust attenuation levels measured in the 3$''$ aperture regions can be applicable to the outer regions. Although this approach might bring some uncertainties, we believe that our simple reproducible approach works reasonably well. We will show a comparison between the SFR$_{\rm H\alpha}$ derived with this approach and the independently measured UV$+$FIR-based SFRs in Section~2.3 (see also \cite{Koyama2015}; \cite{Shimakawa2017}; \cite{Matsuki2017}), and we will also discuss the potential effects of aperture correction in Section~3.4.

\begin{figure*}
\vspace{-4mm}
 \begin{center}
  \includegraphics[width=17.0cm,angle=0]{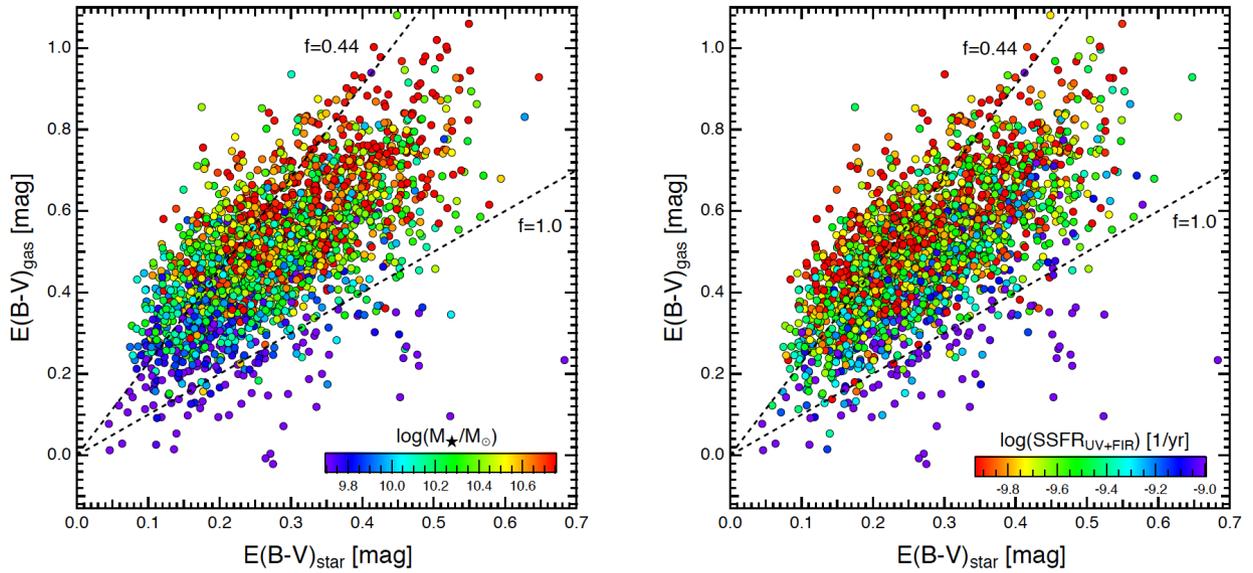} 
 \end{center}
\vspace{-3mm}
\caption{A comparison between dust reddening to nebular emission (E(B$-$V)$_{\rm gas}$) and stellar light (E(B$-$V)$_{\rm star}$) for our AKARI--SDSS--GALEX sample. The color coding indicates stellar mass and specific SFR of individual galaxies in the left and right panel, respectively; i.e.\ redder symbols show more massive galaxies in the left panel, whilst redder symbols indicate lower specific SFR in the right panel. We find a trend that galaxies with higher stellar mass and/or lower specific SFR tend to be located at the upper-side envelope of this correlation---i.e.\ galaxies with higher stellar mass and/or lower specific SFR tend to have higher ``extra'' attenuation toward nebular regions. The dashed lines drawn in each plot show the relation of $f$$=$E(B$-$V)$_{\rm star}$/E(B$-$V)$_{\rm star}$$=$0.44 and 1.0 for reference.}\label{fig:EBVstar_EBVgas}
\end{figure*}

\subsection{UV- and FIR-based SFRs and dust attenuation to stellar light}

We derive dust-unobscured part of SFRs from {\it GALEX} FUV ($\lambda_{\rm eff}$$=$1516~\AA) photometry using \citet{Kennicutt1998} calibration assuming the \citet{Kroupa2001} IMF: 
\begin{equation}
{\rm SFR}_{\rm UV} = 9.7 \times 10^{-29}L_{\nu,{\rm FUV}} {\rm [erg\,s^{-1}\,Hz^{-1}]}.
\end{equation} 
We use ``{\sc fuv\_mag\_best}'' in the \citet{Bianchi2011} catalog as the total FUV magnitudes, and apply the foreground Galactic extinction correction using the \citet{Schlegel1998} dust map and the Galactic extinction curve of \citet{Cardelli1989} with $R_V=3.1$. We also apply the $k$-correction by using a $k$-correction tool developed by \citet{Chilingarian2010} and \citet{Chilingarian2012}, where the required correction values are predicted for individual galaxies based on their FUV$-$NUV and NUV$-$$g$ colors. We note that the $k$-correction values applied to the FUV magnitude of our sample are very small (typically $\sim$0.02-mag).

We also derive the dust-obscured part of SFRs from {\it AKARI} FIR data. Following \citet{Koyama2015}, we derive the total infrared luminosity ($L_{\rm FIR}$) using the WIDE-S (90$\mu$m) and WIDE-L (140$\mu$m) photometry with the equation established by \citet{Takeuchi2010}: 
\begin{equation}
\log L_{\rm FIR} = 0.964\log L^{\rm 2band}_{\rm AKARI} + 0.814,
\end{equation}
where $L^{\rm 2band}_{\rm AKARI} = \Delta \nu_{\rm 90\mu m} L_{\nu}({\rm 90\mu m}) + \Delta \nu_{\rm 140\mu m} L_{\nu}({\rm 140\mu m})$. We note that $\Delta \nu _{\rm 90\mu m}$ ($=$1.47$\times 10^{12}$[Hz]) and $\Delta \nu _{\rm 140\mu m}$ ($=$0.831$\times 10^{12}$[Hz]) denote the band widths of AKARI WIDE-S and WIDE-L bands, respectively. We realize that 187 out of 2,416 galaxies (7.7\%) are detected only at WIDE-S (90$\mu$m) band, and we remove these sources from our sample (because we cannot estimate their $L_{\rm FIR}$ with the above technique). We do not consider $k$-correction for the FIR photometry as the AKARI band widths are wide enough and its effect is negligible for our $z<0.1$ galaxy sample. We then use \citet{Kennicutt1998} equation with the \citet{Kroupa2001} IMF to derive IR-based SFR: 
\begin{equation}
{\rm SFR}_{\rm FIR} = 3.1 \times 10^{-44} (1-\eta) L_{\rm IR} {\rm [erg \cdot s^{-1}]}, 
\end{equation}
where $\eta$ indicates the fraction of IR emission heated by old stars. We here adopt a constant $\eta$$=$0.17 for all our AKARI-detected sources, following the average $\eta$ value for local star-forming galaxies reported by \citet{Buat2011}. We note that the $\eta$ value could be dependent on $L_{\rm IR}$ or specific SFR of individual galaxies, while the reported $\eta$ values are different from studies to studies within a range of $\eta$$=$0.1--0.4 (e.g.\ \cite{Hirashita2003}; \cite{Bell2003}; \cite{Buat2011}). We stress that our conclusions are not affected even if we choose different $\eta$ values. Fig.~\ref{fig:SFRHa_SFRFIR_compare} compares SFR$_{\rm UV + FIR}$ and the dust-corrected SFR$_{\rm H\alpha}$. Although there exists a relatively large scatter ($\sigma$$=$0.31~dex), these two independently measured SFRs show good agreement with each other (with a median difference of $\sim$0.05~dex). We note that we here further removed 101 sources (4.5\%) having negative aperture correction in the SDSS $r$-band data (with $m_{r,{\rm fiber}}$$<$$m_{r, {\rm petro}}$), as these galaxies show unrealistically small SFR$_{\rm H\alpha}$.

Finally, we derive dust attenuation at FUV ($A_{\rm FUV}$) as well as the color excess toward stellar light (E(B$-$V)$_{\rm star}$) by comparing the total (UV$+$FIR) SFRs and the observed UV-derived SFRs (without dust attenuation correction): 
\begin{equation}
A_{\rm FUV} = 2.5 \times \log ({\rm SFR_{UV + FIR}/SFR_{UV}}),
\end{equation}
\begin{equation}
E(B-V)_{\rm star} = A_{\rm FUV} / k_{\rm FUV},
\end{equation}
where we determine $k_{\rm FUV}$ using the \citet{Calzetti2000} reddening curve. In summary, our final {\it AKARI--SDSS--GALEX} star-forming galaxy sample includes 2,128 sources with the measured SFRs and dust attenuation. The source catalog as well as our measurements of SFR and dust attenuation for individual sources will be published on the {\it AKARI} website.

\begin{figure*}
 \vspace{-5mm}
 \begin{center}
  \includegraphics[width=17.5cm,angle=0]{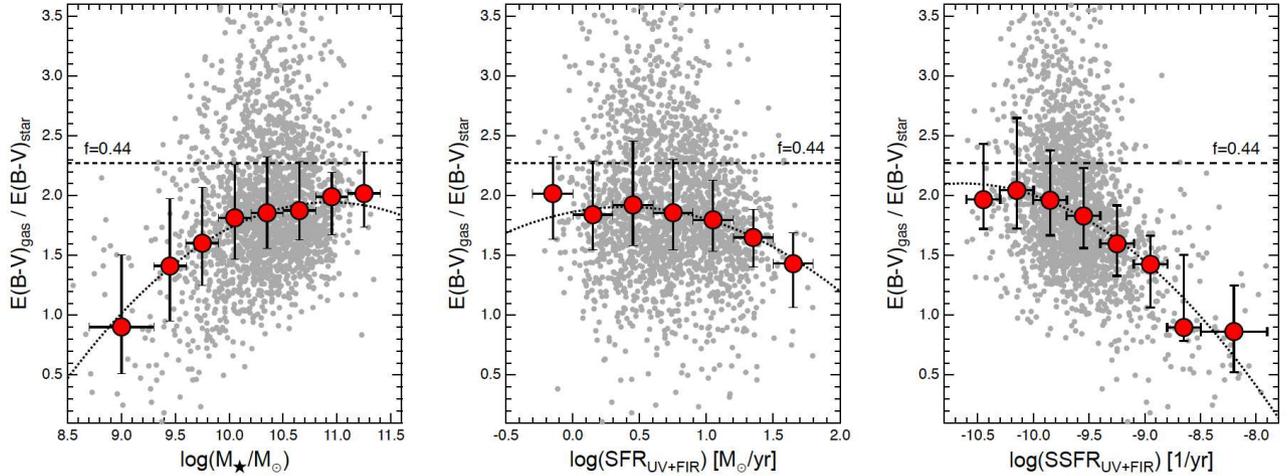} 
 \end{center}
\vspace{-3mm}
\caption{The extra attenuation toward nebular emission quantified by E(B$-$V)$_{\rm gas}$/E(B$-$V)$_{\rm star}$ ratio, plotted against stellar mass (left), SFR (middle), and specific SFR (right) for our AKARI--SDSS--GALEX sample. The gray circles indicate individual AKARI--SDSS--GALEX galaxies, and the red circles with error-bars show their median data points computed for a constant 0.3-dex bin, with the y-axis error-bars representing their 25\%--75\% distribution. We note that a wider bin size is adopted at the lowest-mass and highest sSFR bins so that each bin has sufficient ($>$15) number of galaxies. The error bars in x-axis indicate the adopted bin size. The dotted-line curve drawn in each panel shows the best-fit polynomial function computed for the median data points weighted by $\sigma$/$\sqrt{N}$ of each bin, and the horizontal dashed line indicates the canonical value of E(B$-$V)$_{\rm star}$/E(B$-$V)$_{\rm gas}$$=$0.44 for reference.}   
\label{fig:extra_vs_properties}
\end{figure*}

\subsection{General properties of AKARI-detected galaxies}
\label{sec:dust_extinction}

Because we aim to study the dust attenuation properties of star-forming galaxies with our new {\it AKARI--SDSS--GALEX} sample, it is important to understand fundamental properties and any potential bias associated with the sample. We here briefly examine the nature of the {\it AKARI--SDSS--GALEX} sample selected in the previous sections. 

In Fig.~\ref{fig:sample_on_SFMS}, we show our new {\it AKARI--SDSS--GALEX} galaxies on the SFR--$M_{\star}$ diagram (red symbols) on top of the distribution of the parent {\it SDSS--GALEX} star-forming galaxy sample (gray-scale image plot). We here use SFR$_{\rm H\alpha,corr}$ for both {\it AKARI--SDSS--GALEX} and {\it SDSS--GALEX} samples (as we cannot estimate SFR$_{\rm FIR}$ for galaxies without {\it AKARI} detection). As expected, the {\it AKARI}-detected sources are relatively massive (mostly with $\log(M_{\star}/M_{\odot}) \gtrsim 10.0$), and highly star-forming population (with SFR$\gtrsim$1~$M_{\odot}$/yr). We note that $\sim$7\% of the {\it AKARI--SDSS--GALEX} galaxies have $>$4$\times$ higher SFRs with respect to the main sequence (hence starbursts), and so the majority of our samples are massive (normal) star-forming galaxies.

We then plot in Fig.~\ref{fig:AHa_vs_properties} the H$\alpha$ dust attenuation ($A_{\rm H\alpha}$) against their stellar mass (left), SFR (middle), and specific SFR (right). Again, our {\it AKARI--SDSS--GALEX} galaxies are shown with the red symbols, while the overall distribution of the {\it SDSS--GALEX} sample is shown with the gray-scale image plot. It seems that the distribution of the {\it AKARI--SDSS--GALEX} galaxies is similar to that of all the {\it SDSS--GALEX} sample on the $A_{\rm H\alpha}$--$M_{\star}$ and $A_{\rm H\alpha}$--SFR diagrams (except that they are skewed to massive/high SFR side), while they tend to show high $A_{\rm H\alpha}$ for their specific SFRs. This is not surprising considering the fact that low-mass galaxies (with $\log(M_{\star}/M_{\odot}) \lesssim 10.0$) are not detected at FIR regardless of their specific SFR (Fig~\ref{fig:sample_on_SFMS}), and those with high specific SFR (but with low $A_{\rm H\alpha}$) are dominated by low-mass galaxies. We should keep this potential bias in mind, but we believe that our conclusions will not be affected as we will show consistent results in Section~5 by using {\it WISE} MIR data which is much deeper in terms of IR-based SFRs.

\section{Different dust attenuation toward stellar and nebular regions with AKARI sample}

In Fig.~\ref{fig:EBVstar_EBVgas}, we compare E(B$-$V)$_{\rm gas}$ and E(B$-$V)$_{\rm star}$ for our {\it AKARI--SDSS--GALEX} sample. It can be seen that there is a positive correlation between E(B$-$V)$_{\rm gas}$ and E(B$-$V)$_{\rm star}$ with a large scatter. In the left and right panel of Fig.~\ref{fig:EBVstar_EBVgas}, the color coding indicates stellar mass and specific SFR of individual galaxies, respectively; i.e.\ redder symbols show more massive galaxies in the left panel, while redder symbols indicate lower sSFR in the right panel. We find that galaxies with higher $M_{\star}$ and/or galaxies with lower specific SFR dominate the upper side of the overall distribution of the data points, suggesting that galaxies with higher $M_{\star}$ and/or lower sSFR tend to have higher levels of ``extra'' attenuation toward nebular regions at fixed E(B$-$V)$_{\rm star}$. We showed a similar plot in \citet{Koyama2015} (see their fig.~14), but we have now confirmed the results using a larger sample size with our updated (deeper) {\it AKARI--SDSS--GALEX} catalog.

We here quantify the levels of extra reddening toward nebular regions by calculating the E(B$-$V)$_{\rm gas}$/E(B$-$V)$_{\rm star}$ ratio, and study the correlation between E(B$-$V)$_{\rm gas}$/E(B$-$V)$_{\rm star}$ and various galaxy properties. In Fig.~\ref{fig:extra_vs_properties}, we plot the E(B$-$V)$_{\rm gas}$/E(B$-$V)$_{\rm star}$ ratio as functions of stellar mass (left), SFR (middle), and specific SFR (right). We also show the running median in each panel; red circles with error-bars representing the 25\%--75\% distribution in each bin. Although the scatter is large in all the panels, the left panel of Fig.~\ref{fig:extra_vs_properties} demonstrates the trend that E(B$-$V)$_{\rm gas}$/E(B$-$V)$_{\rm star}$ increases with increasing stellar mass, consistent with the visual impression that we recognized in Fig.~\ref{fig:EBVstar_EBVgas}. We therefore conclude that more massive galaxies tend to have higher levels of extra attenuation toward nebular regions, and the best-fit relation between E(B$-$V)$_{\rm gas}$/E(B$-$V)$_{\rm star}$ and the stellar mass of galaxies is provided with the following equation: 
\begin{equation}
{\rm E(B}-{\rm V)}_{\rm gas}/{\rm E(B}-{\rm V)}_{\rm star} = a_0 + a_1 X + a_2 X^2,
\end{equation}
where $X$$=$$\log(M_{\star}/10^{10}M_{\odot})$, with the best-fit parameters of $a_0$$=$1.729$\pm$0.019, $a_1$$=$0.464$\pm$0.039, and $a_2$$=$$-$0.249$\pm$0.048. 

The middle and right panels of Fig.~\ref{fig:extra_vs_properties} suggest that E(B$-$V)$_{\rm gas}$/E(B$-$V)$_{\rm star}$ mildly decreases with increasing SFR, and as a combination of the trends with stellar mass and SFR, we find that E(B$-$V)$_{\rm gas}$/E(B$-$V)$_{\rm star}$ sharply decreases with increasing specific SFR, again consistent with the results from Fig.~\ref{fig:EBVstar_EBVgas}. Our data suggests that more actively star-forming galaxies tend to have lower E(B$-$V)$_{\rm gas}$/E(B$-$V)$_{\rm star}$ ratio, and E(B$-$V)$_{\rm gas}$/E(B$-$V)$_{\rm star}$ becomes consistent with unity at the highest sSFR end; i.e.\ no extra attenuation is required for very actively star-forming population. By fitting the data points of the running median in the middle and right panels of Fig.~\ref{fig:extra_vs_properties}, the best-fit relation between E(B$-$V)$_{\rm gas}$/E(B$-$V)$_{\rm star}$ and SFR ($Y$$=$$\log$(SFR) [$M_{\odot}$/yr]) and specific SFR ($Z$$=$$\log$(sSFR)~[1/Gyr]) are provided as follows: 
\begin{equation}
{\rm E(B}-{\rm V)}_{\rm gas}/{\rm E(B}-{\rm V)}_{\rm star} = b_0 + b_1 Y + b_2 Y^2,
\end{equation}
\begin{equation}
{\rm E(B}-{\rm V)}_{\rm gas}/{\rm E(B}-{\rm V)}_{\rm star} = c_0 + c_1 Z + c_2 Z^2,
\end{equation}
with the best-fit parameters of $b_0$$=$1.865$\pm$0.049, $b_1$$=$0.213$\pm$0.129, $b_2$$=$$-$0.273$\pm$0.079, $c_0$$=$1.450$\pm$0.022, $c_1$$=$$-$0.798$\pm$0.050, and $c_2$$=$$-$0.244$\pm$0.048.
We believe that the above equations can be a convenient tool to provide a realistic estimate of the extra attenuation toward nebular regions (E(B$-$V)$_{\rm gas}$/E(B$-$V)$_{\rm star}$), but we caution that the equations should be used with care, in particular when one needs to use the equation outside the $M_{\star}$/SFR/sSFR range shown in Fig.~\ref{fig:extra_vs_properties}. For example, the equations provide unrealistic ``negative'' extra attenuation toward nebular regions (i.e.\ E(B$-$V)$_{\rm gas}$/E(B$-$V)$_{\rm star}$$<$1) at the very low-mass end as well as at high specific SFR ends. It would be more realistic to assume E(B$-$V)$_{\rm gas}$/E(B$-$V)$_{\rm star}$$\sim$1 (hence no extra attenuation to nebular regions) at the low-mass or high-SSFR end.

As discussed in Section~1, some recent high-redshift studies suggest that the extra attenuation toward nebular regions for high-$z$ galaxies are different from that of local galaxies (e.g.\ \cite{Erb2006}; \cite{Reddy2010}; \cite{Kashino2013}; \cite{Pannella2015}; \cite{Forster-Schreiber2009}; \cite{Tadaki2013}; \cite{Wuyts2013}; \cite{DeBarros2016}; see \cite{Puglisi2016} for review), but the derived E(B$-$V)$_{\rm gas}$/E(B$-$V)$_{\rm star}$ values are different from studies to studies most likely depending on how the parent samples are selected. Our results suggest that the extra attenuation toward nebular regions with respect to that to stellar light significantly changes with stellar mass and/or specific SFR even for the local galaxies, and it is not surprising to see the extra attenuation levels are different for high-$z$ galaxies because typical SF activity of high-$z$ star-forming galaxies is an order of magnitude higher than those in the present-day universe (e.g.\ \cite{Madau2014}; \cite{Daddi2007}; \cite{Elbaz2007}; \cite{Whitaker2012}). 

\begin{figure}
\vspace{-3mm}
 \begin{center}
\includegraphics[width=8.1cm,angle=0]{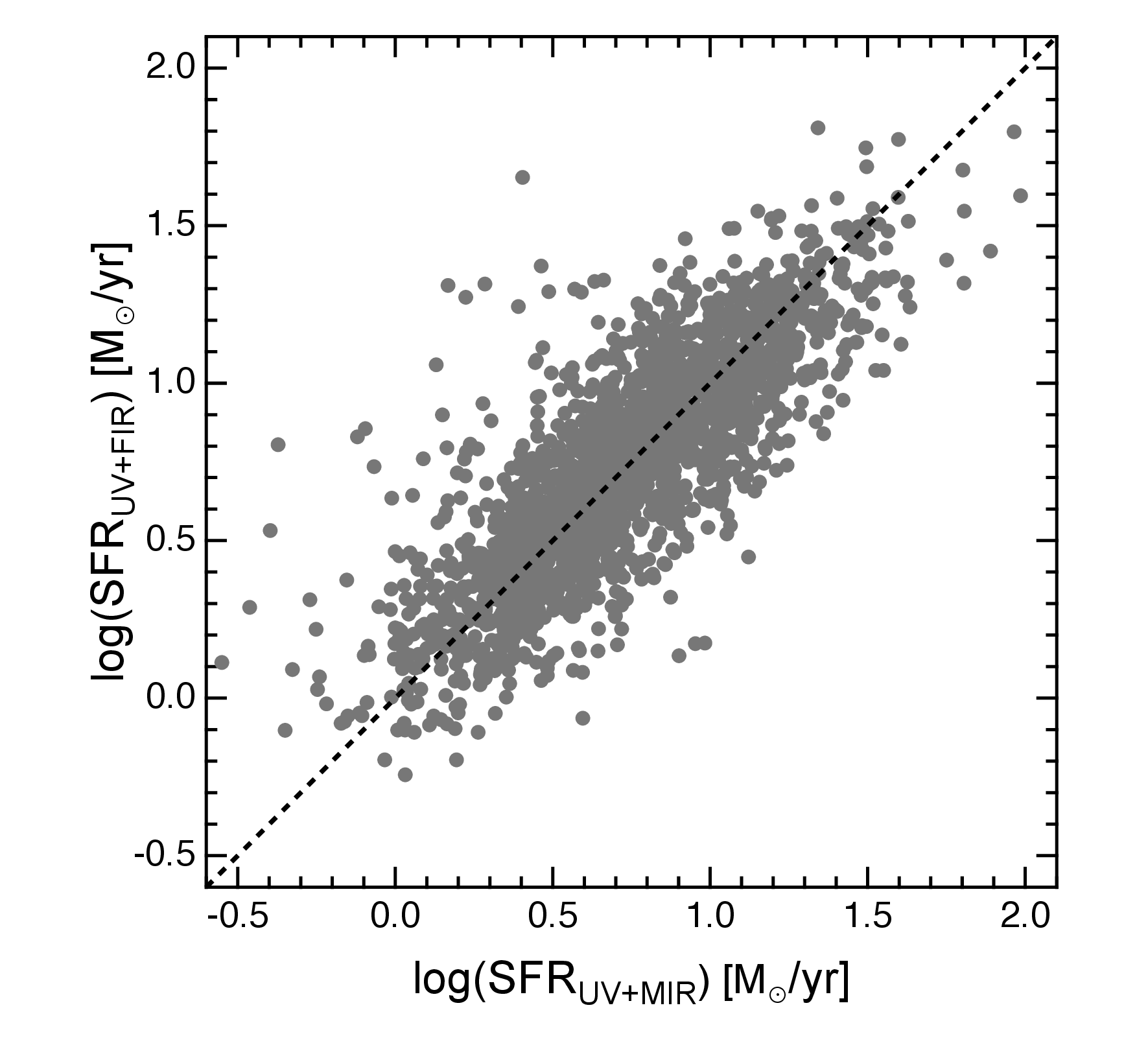} 
 \end{center}
\vspace{-4mm}
\caption{SFRs derived from AKARI FIR photometry (Section~2.2) plotted against SFRs from WISE 22$\mu$m data taken from the public GSWLC catalog by \citet{Salim2016} for galaxies identified in both our AKARI--SDSS--GALEX sample and the GSWLC catalog, taking into account the different IMF assumption. The two SFR measurements show good agreement with each other. }
\label{fig:SFR_compare_WISE}
\end{figure}
\begin{figure*}
\vspace{-5mm}
 \begin{center}
 \includegraphics[width=17.5cm,angle=0]{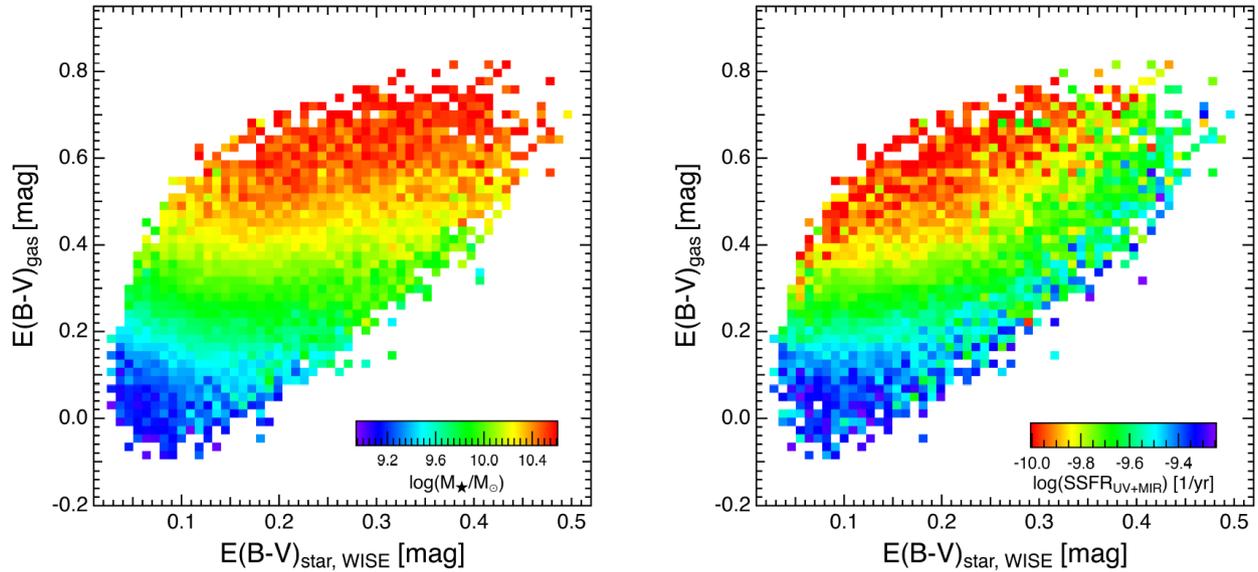}
 \end{center}
\vspace{-2mm}
\caption{E(B$-$V)$_{\rm gas}$ versus E(B$-$V)$_{\rm star}$ plot for SDSS--GALEX--WISE matched sample, with color coding based on stellar mass (left) and specific SFR (right). To create this plot, we divide the panels into 60$\times$60 sub-grids and compute the average stellar mass and sSFR at each pixel. We request a minimum sample size of $N_{\rm galaxy}$$=$4 to compute the average stellar mass in this diagram. We follow this strategy when we create similar color-image plots in the following sections. This plot confirms the trend we reported in Fig.~\ref{fig:EBVstar_EBVgas} with larger sample; i.e.\ galaxies with higher $M_{\star}$ and/or with lower specific SFR tend to have higher E(B$-$V)$_{\rm gas}$ at fixed E(B$-$V)$_{\rm star}$. }
\label{fig:EBVstar_EBVgas_WISE}
\end{figure*}

\section{A sanity check with WISE MIR photometry}
\label{sec:WISE}

The main goal of this paper is to use the newly released {\it AKARI} Faint Source Catalog ver.1 and show its scientific capability. It is therefore important to test the validity of our scientific results presented in this work with some independent dataset. The Wide-field Infrared Survey Explorer ({\it WISE}; \cite{Wright2010}) all-sky MIR survey data can be an ideal tool for this purpose, because it can provide independent measurements of IR-based SFRs. We here use the {\it GALEX-SDSS-WISE} Legacy Catalog (GSWLC) published by \citet{Salim2016}\footnote{https://archive.stsci.edu/prepds/gswlc/}, in which they provide WISE 22$\mu$m-based SFRs (SFR$_{\rm MIR}$) for all galaxies detected at 22$\mu$m, by interpolating the \citet{Chary2001} SEDs to match the observed 22$\mu$m flux densities. In this work, we use the SFR$_{\rm MIR}$ from 22$\mu$m flux in ``unWISE'' catalog (\cite{Lang2016}), recommended for low-$z$ galaxies with large apparent size (see \cite{Salim2016}). We note that \citet{Salim2016} assume \citet{Chabrier2003} IMF when deriving the SFR$_{\rm MIR}$, and we therefore rescaled their SFR$_{\rm MIR}$ (by $+$0.03~dex) to account for the IMF difference.

Following the same strategy as that we performed in Section~2.1 to construct our {\it AKARI--SDSS--GALEX} catalog, we first match our {\it SDSS--GALEX} star-forming galaxy sample (at 0.02$<$$z$$<$0.1) with the ``GSWLC-X'' catalog\footnote{We note that \citet{Salim2016} published three versions of the catalog (GSWLC-A, M, D) depending on the {\it GALEX} survey depths used for the source matching, and they also publish a master catalog (GSWLC-X), in which they exploit the deepest data among GSWLC-A, M, and D. We decided to use this ``GSWLC-X'' which covers the largest fraction of SDSS survey area (see \cite{Salim2016} for more details).}. We here reconstructed {\it WISE--SDSS--GALEX} catalog by matching our own {\it SDSS--GALEX} sample with the GSWLC catalog, to keep consistency in the data release versions of the {\it SDSS} and {\it GALEX} catalog used for the analyses. We restrict the sample to those detected at WISE 22$\mu$m band, and now we identify 50,353 {\it WISE--SDSS--GALEX} sample for which MIR-based SFRs are available. By matching the {\it WISE--SDSS--GALEX} sample with our {\it AKARI--SDSS--GALEX} sample constructed in Section~2.1, we find that 2,036 sources ($>$95\%) of the {\it AKARI--SDSS--GALEX} sample are also identified in the {\it WISE--SDSS--GALEX} sample constructed here. 

In Fig.~\ref{fig:SFR_compare_WISE}, we compare the FIR-based SFRs (derived with {\it AKARI} data in Section~2.2) and {\it WISE} 22$\mu$m-based SFRs (provided by \cite{Salim2016}, after accounting for the IMF difference) for galaxies identified in both {\it AKARI} and {\it WISE} catalogs, showing a reasonable agreement between UV$+$FIR and UV$+$MIR SFRs (with a median difference of 0.01~dex and a scatter of $\sigma$$=$0.22~dex). 
Following the good agreement between the FIR- and MIR-based SFRs, we will perform the same analysis presented in the previous sections with a significantly larger number of galaxies in the {\it WISE--SDSS--GALEX} sample. 

We here apply the same method as we performed in Section~2.2 to derive E(B$-$V)$_{\rm star}$ from SFR$_{\rm FIR}$ and SFR$_{\rm UV}$, and we now calculate $A_{\rm UV,WISE}$ and E(B$-$V)$_{\rm star,WISE}$ with the following equations: 
\begin{equation}
A_{\rm FUV,WISE} = 2.5 \times \log ( {\rm SFR_{UV+MIR}} / {\rm SFR_{UV}} ),
\end{equation}
\begin{equation}
E(B-V)_{\rm star,WISE} = A_{\rm FUV, WISE} / k_{\rm FUV}.
\end{equation}
In Fig.~\ref{fig:EBVstar_EBVgas_WISE}, we compare E(B$-$V)$_{\rm gas}$ and E(B$-$V)$_{\rm star,WISE}$, with color coding based on stellar mass and specific SFR of galaxies. This is the same plot as Fig.~\ref{fig:EBVstar_EBVgas}, but considering the large sample size of the {\it WISE--SDSS--GALEX} sources, we here show the average $M_{\star}$ (left panel) and specific SFR (right panel) at each position on this diagram, rather than showing the individual data points. This plot clearly demonstrates that galaxies with higher stellar mass or lower specific SFR tend to have higher E(B$-$V)$_{\rm gas}$ at fixed E(B$-$V)$_{\rm star}$, confirming the trend we reported in Fig.~\ref{fig:EBVstar_EBVgas} by using {\it AKARI--SDSS--GALEX} sample.

\begin{figure*}
 \vspace{-4mm}
 \begin{center}
  \includegraphics[width=17.5cm,angle=0]{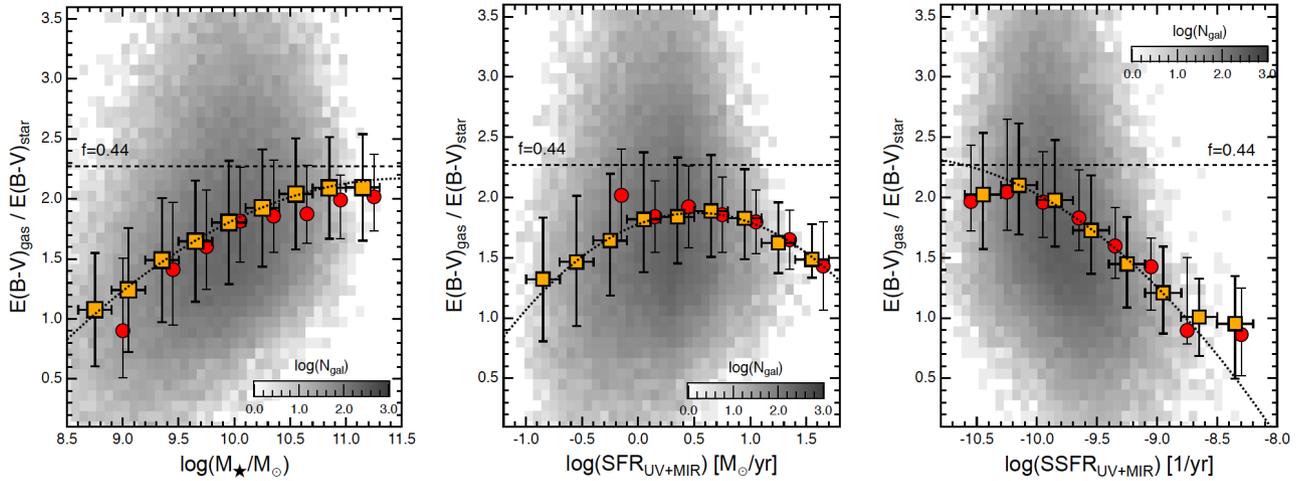} 
 \end{center}
\vspace{-3mm}
\caption{The same plot as Fig.~6 to show the extra attenuation levels as a function of various properties for the SDSS--GALEX--WISE sample. Because of the very large number of data points, we show their distribution by gray-scale density plot, instead of showing the individual data points on these plots. The orange squares with error-bars show the running median (with 25\%--75\% distribution), and the best-fit 3rd-order polynomial function curve is also shown in each panel. The horizontal error bars represent the 0.3-dex bin size applied for all the calculation in this diagram, where all the subsamples have a sufficient number of galaxies ($>$50). The median data points plotted on Fig.~\ref{fig:extra_vs_properties} (derived with AKARI data) are also shown with the red circles, showing good agreement with the results using {\it WISE} data, and the canonical E(B$-$V)$_{\rm star}$/E(B$-$V)$_{\rm gas}$$=$0.44 line is drawn for reference.}\label{fig:extra_vs_properties_WISE}
\end{figure*}

We then quantify the extra reddening toward nebular regions by computing the E(B$-$V)$_{\rm gas}$/E(B$-$V)$_{\rm star,WISE}$ ratio (as we did in the previous section), and we plot in Fig.~\ref{fig:extra_vs_properties_WISE} the derived E(B$-$V)$_{\rm gas}$/E(B$-$V)$_{\rm star,WISE}$ against their stellar mass (left), SFR (middle), and specific SFR (right) for all {\it WISE--SDSS--GALEX} galaxies. Instead of showing the individual data points, we show the distribution of all {\it WISE--SDSS--GALEX} galaxies with the gray-scale density plot, and we show the running median with the orange squares with error-bars (indicating the 25\%--75\% distribution in each bin). It can be seen that the extra attenuation level increases with increasing stellar mass, while it decreases with increasing specific SFR, consistent with the results we reported in Fig.~\ref{fig:extra_vs_properties}.

The trend for SFR seems to be different from that shown in Fig.~\ref{fig:extra_vs_properties}; we showed a mild {\it decrease} of the E(B$-$V)$_{\rm gas}$/E(B$-$V)$_{\rm star}$ ratio with increasing SFR in Fig.~\ref{fig:extra_vs_properties}, while in Fig.~\ref{fig:extra_vs_properties_WISE}, E(B$-$V)$_{\rm gas}$/E(B$-$V)$_{\rm star}$ increases at $\log$(SFR)$\lesssim$$-$1.0--0.5 [M$_{\odot}$/yr] and decreases at $\log$(SFR)$\gtrsim$0.5--1.5~[M$_{\odot}$/yr]. However, it should be noted that the SFR range presented in Fig.~\ref{fig:extra_vs_properties} and Fig.~\ref{fig:extra_vs_properties_WISE} are different. As we showed in Section~3.1, most of the {\it AKARI}-detected samples have $\log$(SFR)$\gtrsim$0~[M$_{\odot}$/yr], whilst the {\it WISE} data goes an order of magnitude deeper than the {\it AKARI} data in terms of IR-based SFR. We note that the mild decrease of the E(B$-$V)$_{\rm gas}$/E(B$-$V)$_{\rm star,WISE}$ ratio toward the high SFR end seen in Fig.~\ref{fig:extra_vs_properties} is consistent with our current analysis using {\it WISE} data (see the red circles in Fig.~\ref{fig:extra_vs_properties_WISE} showing the same data points as in Fig.~\ref{fig:extra_vs_properties}). The reason of this different behavior at low-SFR and high-SFR side is unclear, but we speculate that the E(B$-$V)$_{\rm gas}$/E(B$-$V)$_{\rm star}$ ratio increases following the increase of stellar mass at low-SFR side, while the effect of specific SFR becomes stronger at high-SFR end (the trend with stellar mass becomes almost flat at high-$M_{\star}$ end as shown in the left panel of Fig~\ref{fig:extra_vs_properties_WISE}). 

\section{Extra attenuation toward nebular regions across star-forming main sequence}
\label{sec:MS}

\begin{figure*}
\vspace{-5mm}
 \begin{center}
\includegraphics[width=17.0cm,angle=0]{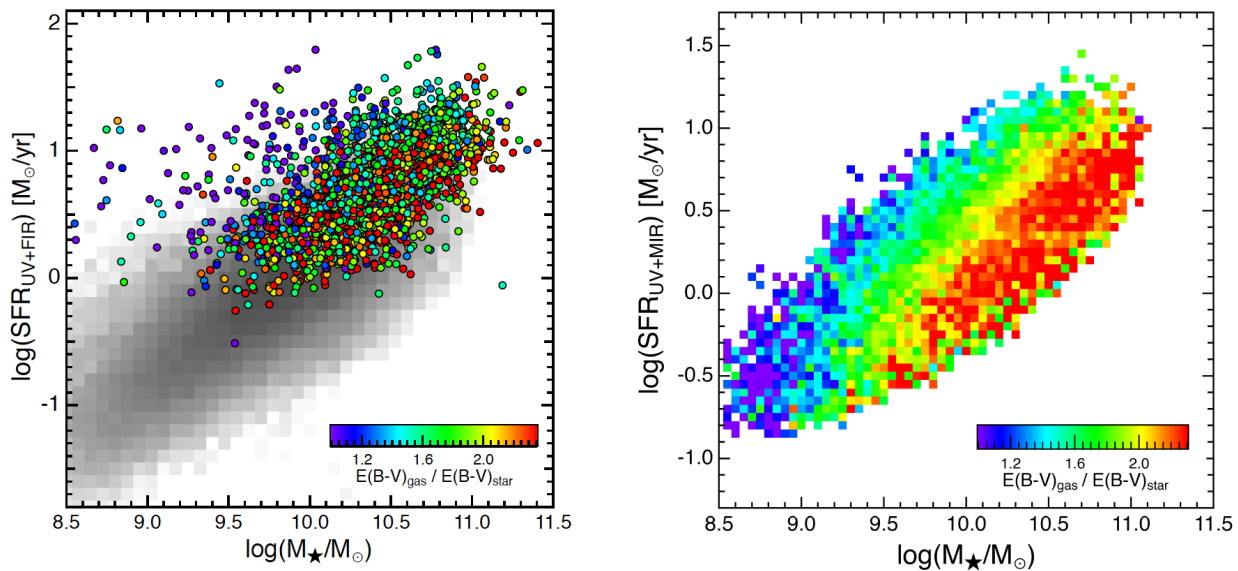} 
 \end{center}
\vspace{-3mm}
\caption{The extra attenuation toward nebular regions across the SFR--$M_{\star}$ diagram for AKARI--SDSS--GALEX sample (left) and for WISE--SDSS--GALEX sample (right). In the left panel, we show the individual data points of the AKARI--SDSS--GALEX sample with color coding indicating E(B$-$V)$_{\rm gas}$/E(B$-$V)$_{\rm star}$ ratio, and the gray-scale density plot shows the distribution of the parent SDSS--GALEX sample (same as Fig.~3). For the right panel, we compute the average E(B$-$V)$_{\rm gas}$/E(B$-$V)$_{\rm star}$ ratio of the WISE--SDSS--GALEX sample at each position on this diagram by dividing each panel into 60$\times$60 sub-grids. We note that we here request a minimum sample size of $N_{\rm galaxy}$$=$4 for each pixel, and thus the data points seen at the top-left corner of the left panel (i.e.\ low-mass galaxies with high SFR) are not seen in the right panel. The two plots are broadly consistent with each other in the sense that the E(B$-$V)$_{\rm gas}$/E(B$-$V)$_{\rm star}$ ratio becomes largest at low-SFR side.}
\label{fig:extra_on_SFMS}
\end{figure*}

We have shown that the extra dust attenuation toward nebular regions changes with stellar mass and (specific) SFR. In this section, we study how the E(B$-$V)$_{\rm gas}$/E(B$-$V)$_{\rm star}$ ratio changes across the SFR--$M_{\star}$ diagram. In Fig.~\ref{fig:extra_on_SFMS}, we show the SFR--$M_{\star}$ diagram for AKARI--SDSS--GALEX sample (left) and for WISE--SDSS--GALEX sample (right) with color coding based on the E(B$-$V)$_{\rm gas}$/E(B$-$V)$_{\rm star}$ ratio. For the right panel of this plot, we divide the diagram into 60$\times$60 sub-grids and compute the average E(B$-$V)$_{\rm gas}$/E(B$-$V)$_{\rm star}$ at each pixel (with requirement of the minimum sample size of $N_{\rm galaxy}$$=$4 in each pixel). Both of these plots demonstrate that, at fixed stellar mass, there remains a trend that galaxies with {\it lower} sSFR tend to have {\it higher} E(B$-$V)$_{\rm gas}$/E(B$-$V)$_{\rm star}$ ratio; i.e.\ galaxies located at the lower side of the star-forming main sequence tend to suffer from higher extra attenuation toward nebular regions.  

Assuming that the levels of extra attenuation toward nebular regions with respect to the stellar continuum light reflects the different geometry of stars and gas/dust within the galaxies as argued by many authors (e.g.\ \cite{Calzetti1997}; \cite{Price2014}; \cite{Reddy2015}), the fact that the E(B$-$V)$_{\rm gas}$/E(B$-$V)$_{\rm star}$ ratio changes across the SFR--$M_{\star}$ diagram suggests that the geometry of stars and dust within the galaxies become more and more different as the SF quenching process proceeds. In other words, the distribution of star-forming regions becomes more distinct (patchy and/or centrally concentrated) from the distribution of stellar components. 

\begin{figure*}
\vspace{-4mm}
 \begin{center}
\includegraphics[width=17.0cm,angle=0]{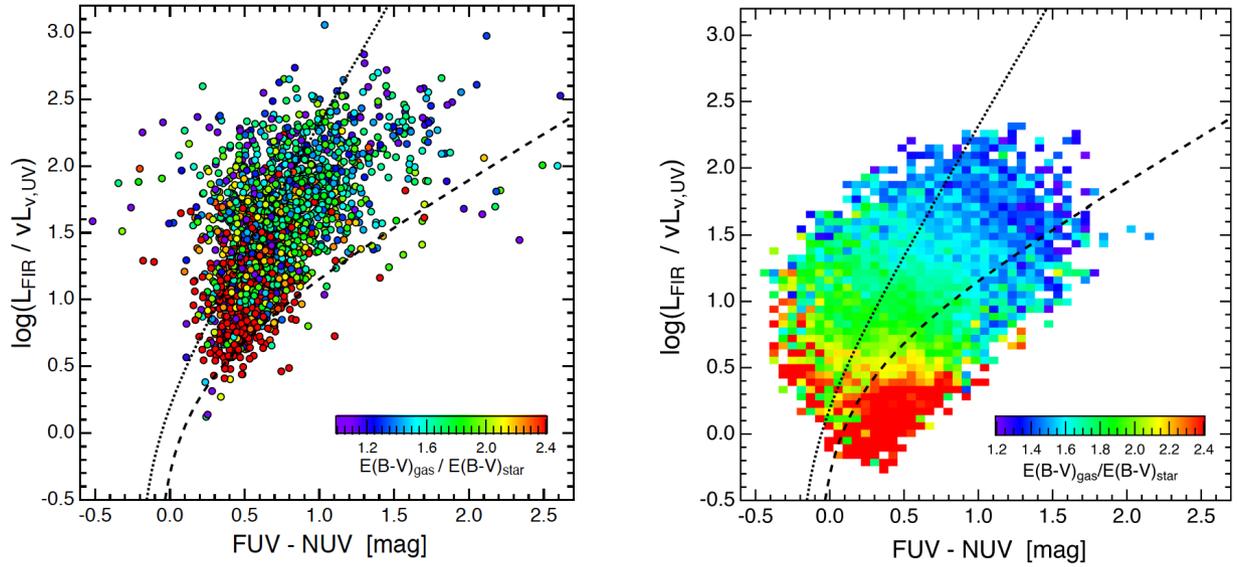}
 \end{center}
\vspace{-3mm}
\caption{The attenuation at FUV (quantified by SFR$_{\rm UV+FIR}$/SFR$_{\rm UV}$ or SFR$_{\rm UV+MIR}$/SFR$_{\rm UV}$ ratio) plotted against FUV$-$NUV colors (corresponding to the UV slope) for the AKARI--SDSS--GALEX sample (left) and WISE--SDSS--GALEX star-forming galaxy sample (right), with the color code indicating the extra reddening toward nebular regions, E(B$-$V)$_{\rm gas}$/E(B$-$V)$_{\rm star}$, in both cases. The dotted-line and dashed-line curves shown in the plots indicate the relation for star-burst galaxies by \citet{Meurer1999} (see also \cite{Takeuchi2010}) and that of more normal star-forming galaxies established by \citet{Boissier2007}, respectively. Dusty star-burst galaxies showing stronger IR excess tend to have {\it lower} extra attenuation toward nebular regions, consistent with the results reported in Fig.~\ref{fig:extra_on_SFMS}. }
\label{fig:IRX_vs_beta}
\end{figure*}

To study the nature of galaxies with high E(B$-$V)$_{\rm gas}$/E(B$-$V)$_{\rm star}$ ratio, we show in Fig.~\ref{fig:IRX_vs_beta} the attenuation at FUV measured with $L_{\rm IR}$/$L_{\rm UV}$ ratio plotted against the FUV$-$NUV colors (equivalent to the so-called ``IRX--$\beta$ diagram; e.g.\ \cite{Meurer1999}; \cite{Witt2000}; \cite{Takeuchi2010}; \cite{Reddy2010}; \cite{Buat2011}), with color coding based on the extra reddening toward nebular regions, E(B$-$V)$_{\rm gas}$/E(B$-$V)$_{\rm star}$. We show individual data points for AKARI--SDSS--GALEX sample, while we show the average of E(B$-$V)$_{\rm gas}$/E(B$-$V)$_{\rm star}$ ratio at each pixel for the WISE--SDSS--GALEX sample. This plot demonstrates that galaxies with ``IR excess'' sources (i.e.\ dusty galaxies) tend to have {\it lower} E(B$-$V)$_{\rm gas}$/E(B$-$V)$_{\rm star}$ ratio, suggesting that the FIR light and the ionized gas emission lines are coming from the same (star-forming) regions within the galaxies. On the other hand, galaxies with high E(B$-$V)$_{\rm gas}$/E(B$-$V)$_{\rm star}$ ratio are strongly clustered at the low $L_{\rm IR}$/$L_{\rm UV}$ side, demonstrating that galaxies with higher E(B$-$V)$_{\rm gas}$/E(B$-$V)$_{\rm star}$ ratio tend to be less dusty, moderately star-forming population, consistent with the results we showed in Fig.~\ref{fig:extra_on_SFMS}. 

\section{Caveats on the use of SFR$_{\rm H\alpha}$ instead of SFR$_{\rm UV+IR}$ and other uncertainties}

\begin{figure}
\vspace{-3mm}
 \begin{center}
\includegraphics[width=7.9cm,angle=0]{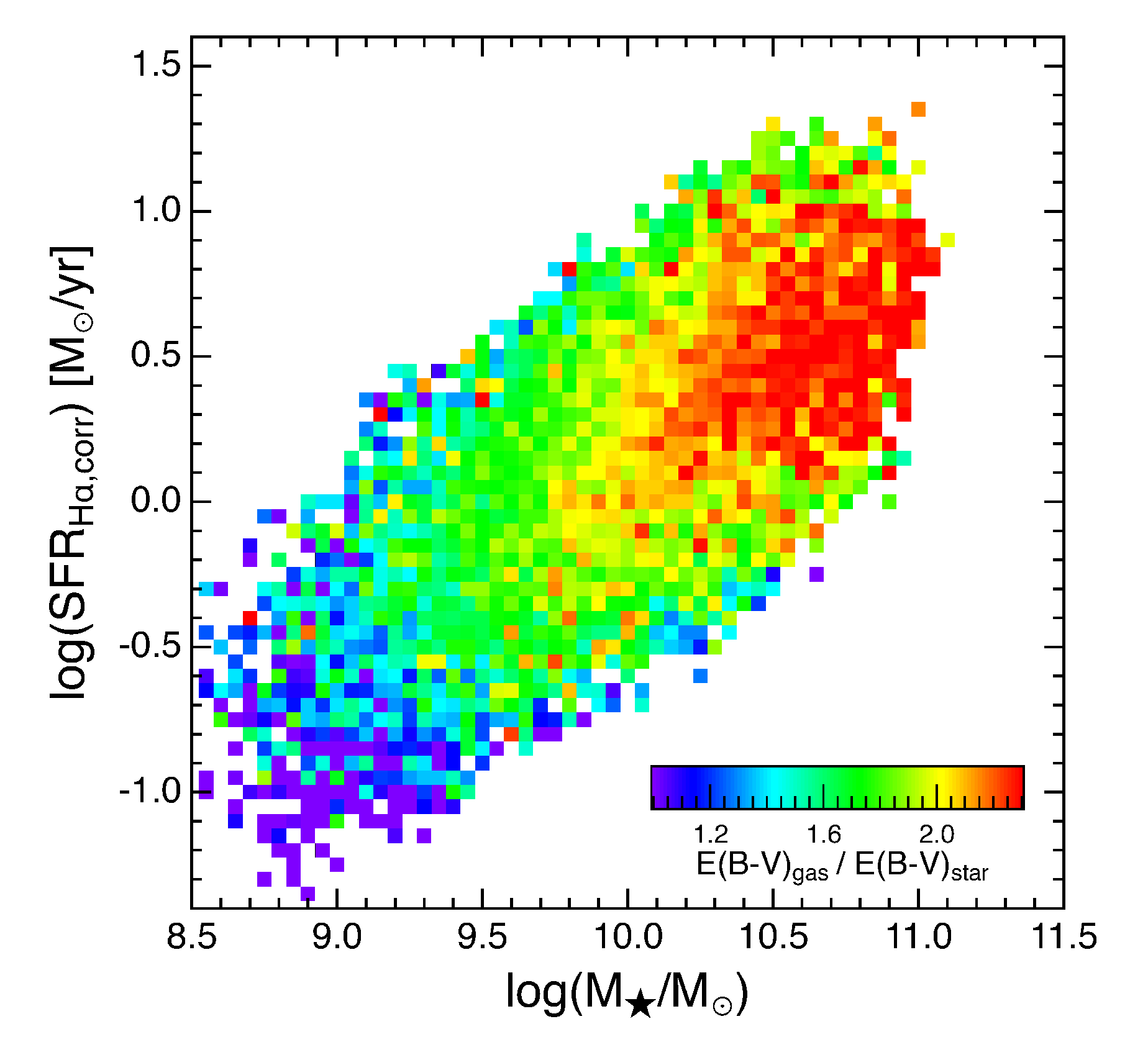} 
 \end{center}
\vspace{-3mm}
\caption{The same plot as Fig.~\ref{fig:extra_on_SFMS}, but here we use H$\alpha$-based SFRs instead of UV$+$IR SFRs. In this plot, it can be seen that the E(B$-$V)$_{\rm gas}$/E(B$-$V)$_{\rm star}$ ratio is higher at the massive end with $\log(M_{\star}/M_{\odot})$$\gtrsim$10.5, while the visual impression on the trend across the SF main sequence is slightly different from that of Fig.~\ref{fig:extra_on_SFMS}, particularly at the low-mass/low-SFR end (see text for more details).}
\label{fig:extra_on_SFMS_Ha}
\end{figure}

\begin{figure}
\vspace{-3mm}
 \begin{center}
\includegraphics[width=8.0cm,angle=0]{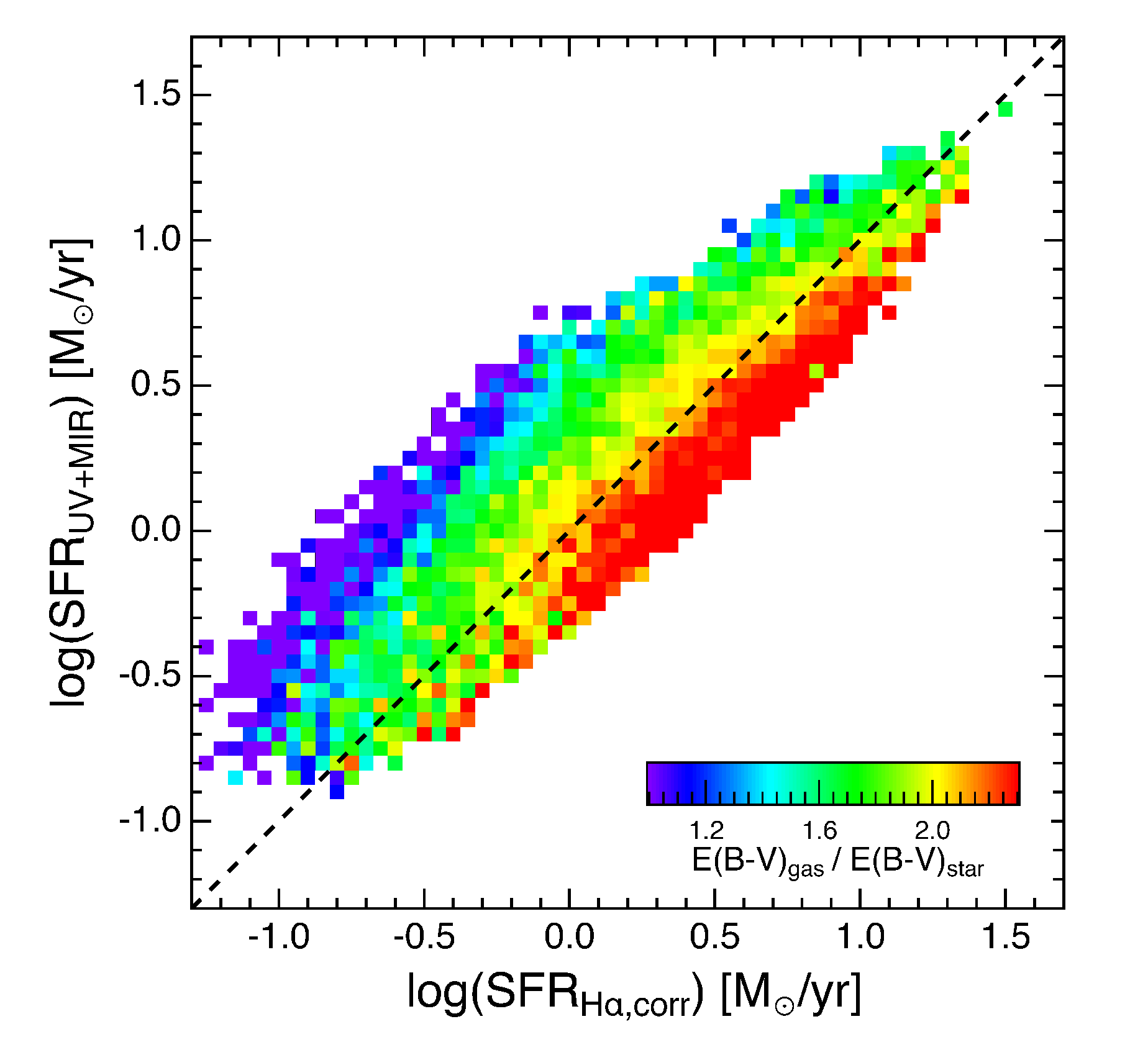} 
 \end{center}
\vspace{-3mm}
\caption{Average extra reddening toward nebular regions, E(B$-$V)$_{\rm gas}$/E(B$-$V)$_{\rm star}$, at each position on the SFR$_{\rm H\alpha}$--SFR$_{\rm UV+MIR}$ diagram. There is a clear trend that galaxies showing SFR$_{\rm H\alpha}$$>$SFR$_{\rm UV+MIR}$ tend to have higher levels of E(B$-$V)$_{\rm gas}$/E(B$-$V)$_{\rm star}$. }
\label{fig:extra_on_SFRSFR}
\end{figure}

In Fig.~\ref{fig:extra_on_SFMS_Ha}, we show the same SFR--$M_{\star}$ plot as Fig.~\ref{fig:extra_on_SFMS} by replacing the SFR$_{\rm UV+FIR}$ (or SFR$_{\rm UV+MIR}$) with the H$\alpha$-derived SFRs. It can be seen that the region showing the highest levels of extra reddening (with E(B$-$V)$_{\rm gas}$/E(B$-$V)$_{\rm star}$$\gtrsim$2; i.e.\ red-color regions in the plot) is strongly concentrated at the very massive end ($\log (M_{\star}/M_{\odot})$$>$10.5), whilst we reported in Fig.~\ref{fig:extra_on_SFMS} that the red-color regions are distributed along the bottom part of the SF main sequence over a wide stellar mass range. The physical reason of the different behavior between SFR$_{\rm UV+MIR}$ and SFR$_{\rm H\alpha,corr}$ is unclear, but we speculate that it is originated from disagreement between the SFR$_{\rm UV+MIR}$ and SFR$_{\rm H\alpha,corr}$ at the low-mass end. We investigate this possibility in Fig.~\ref{fig:extra_on_SFRSFR} by comparing the SFR$_{\rm H\alpha,corr}$ and SFR$_{\rm UV+MIR}$, with the color code indicating the E(B$-$V)$_{\rm gas}$/E(B$-$V)$_{\rm star}$ ratio. It can be seen that galaxies with SFR$_{\rm H\alpha,corr}$$>$SFR$_{\rm UV+MIR}$ tend to show significantly higher levels of extra attenuation. 

\begin{figure*}
\vspace{-5mm}
 \begin{center}
  \includegraphics[width=17.5cm,angle=0]{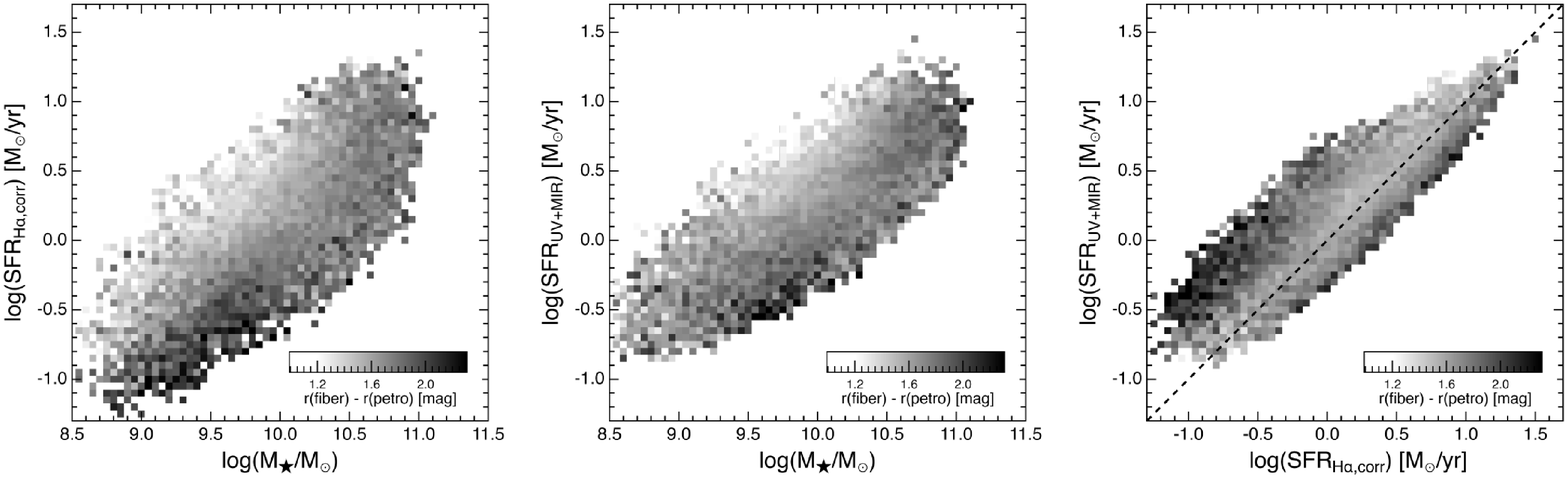} 
 \end{center}
\vspace{-3mm}
\caption{The same plot as Figs.~\ref{fig:extra_on_SFMS}, \ref{fig:extra_on_SFMS_Ha} and \ref{fig:extra_on_SFRSFR}, but here the color coding (gray scale) indicates the average aperture correction ($r_{\rm fiber}$$-$$r_{\rm petro}$) applied for galaxies in each pixel, to examine the potential effects of aperture correction. From the left and middle panels, it can be seen that low-mass galaxies with low SFR tend to require relatively large aperture correction, hence potentially accompanying a large uncertainty. In the right panel, it can be recognized that galaxies showing strong IR excess (with SFR$_{\rm UV+MIR}$$>$SFR$_{\rm H\alpha,corr}$) also tend to require larger aperture correction.}\label{fig:aperture_corr}
\end{figure*}

We caution that the major source of uncertainties of our analysis would be the dust attenuation correction (and aperture correction) when we derive H$\alpha$-based SFRs. We performed a simple dust attenuation correction assuming that the distribution of H$\alpha$ and H$\beta$ line fluxes follow that of stellar continuum light measured in $r$-band (Section~2.2). In Fig.~\ref{fig:aperture_corr}, we briefly test the effects of aperture correction, by creating the same plots as Fig.~\ref{fig:extra_on_SFMS}, Fig.~\ref{fig:extra_on_SFMS_Ha}, and Fig.~\ref{fig:extra_on_SFRSFR} with the color code (gray scale) indicating the average aperture correction at each position on these diagrams ($m_{\rm r,fiber}$$-$$m_{\rm r,petro}$). It can be seen that low-mass galaxies ($\log(M_{\star}/M_{\odot})\lesssim$10) with low SFR tend to require larger aperture correction, mainly because they are intrinsically faint sources and we can detect those low-mass, low-SFR galaxies only when they are located in the very nearby universe (hence with large apparent size). It should be noted that we need to apply $\sim$2-mag or more aperture correction when we derive SFR$_{\rm H\alpha}$ for these low-mass, low-SFR galaxies. We note that the amount of aperture correction is not directly linked to the uncertainties of the derived SFR, but it might be unrealistic to assume the completely uniform dust attenuation levels over the galaxies (e.g. \cite{Kreckel2013}; \cite{Hemmati2015}; \cite{Nelson2016}; \cite{Tacchella2018}), and therefore our results at the low-mass/low-SFR end should be interpreted with care. 



Finally, we comment that another source of uncertainty is regarding the choice of dust attenuation curves. In particular, we should note that the results must be interpreted with care in case one applies variable attenuation curve depending on galaxy properties (e.g.\ \cite{Kriek2013}; \cite{Salim2018}). A recent study by \citet{Salim2018} suggests that lower-mass galaxies or galaxies with low specific SFR tend to have steeper attenuation curve slopes. For those galaxies with steeper slopes, the observed UV attenuation tend to be translated into smaller E(B$-$V)$_{\rm star}$ (hence higher E(B$-$V)$_{\rm gas}$/E(B$-$V)$_{\rm star}$ ratio). This is qualitatively consistent with our results, and may partly explain the trend we reported in this work that the E(B$-$V)$_{\rm gas}$/E(B$-$V)$_{\rm star}$ ratio is higher for galaxies below the SF main sequence. We believe that our conclusions of this work are robust because most of the results are unchanged even if we use SFR$_{\rm H\alpha}$ or SFR$_{\rm UV+IR}$, but a more detailed, systematic spatially-resolved studies for a statistical sample of galaxies selected from above/below the SF main sequence would be needed to obtain a more conclusive evidence for the changing star/dust geometry as well as the changing dust attenuation curve across the SFR--$M_{\star}$ diagram.

\section{Summary}
\label{sec:summary}

In this study, we constructed an updated version of our {\it AKARI--SDSS--GALEX} star-forming galaxy catalog at 0.02$<$$z$$<$0.10 using the newly released AKARI/FIS all-sky Faint Source Catalog ver.1 (FSCv1). With the improved sensitivity of the Faint Source Catalog, the number of the matched sources is increased by a factor of $\sim$2$\times$ compared to the previous version of the catalog presented by \citet{Koyama2015} using the AKARI Bright Source Catalog (ver.1). 

We derived SFRs from dust-corrected H$\alpha$ luminosities and from UV$+$FIR luminosities, and find a reasonable agreement between the two independent measurements of SFRs. We then derived the dust attenuation levels toward stellar light (from IR/UV ratio; SFR$_{\rm UV+FIR}$/SFR$_{\rm UV}$) and nebular emission lines (from H$\alpha$/H$\beta$ ratio), to discuss the different dust attenuation levels toward stellar and nebular regions within local star-forming galaxies. We quantify the ``extra'' attenuation toward nebular regions by E(B$-$V)$_{\rm gas}$/E(B$-$V)$_{\rm star}$ ratio, and find that the levels of extra attenuation increases with increasing stellar mass, while it decreases with increasing specific SFR. At the high specific SFR end, no extra reddening is required (i.e.\ E(B$-$V)$_{\rm gas}$/E(B$-$V)$_{\rm star}$$\approx$1). 

If we assume that the different levels of dust attenuation toward stellar and nebular regions is attributed to the different star/dust geometry within the galaxies as often argued by previous studies, we suggest that more massive galaxies tend to have more patchy (or centrally concentrated) distribution of dust within the galaxies, while the H{\sc ii} regions tend to be more uniformly distributed over the galaxies in low-mass galaxies (or those with high specific SFRs). Our results are also confirmed by using the {\it WISE} 22$\mu$m photometry, with the GSWLC catalog published by \citet{Salim2016}. 

We then study how the extra attenuation toward nebular regions, E(B$-$V)$_{\rm gas}$/E(B$-$V)$_{\rm star}$, changes on the SFR--$M_{\star}$ diagram, by using the {\it AKARI}--{\it SDSS}--{\it GALEX} and  {\it WISE}--{\it SDSS}--{\it GALEX} samples. We find that, even at a fixed stellar mass, there is a clear trend that galaxies located {\it below} the star-forming main sequence tend to have higher levels of extra attenuation, suggesting that the change in the E(B$-$V)$_{\rm gas}$/E(B$-$V)$_{\rm star}$ ratio is related to the galaxy transition or SF quenching process, and that the distribution of stars and dust becomes more and more different as the SF quenching process proceeds.

Our results based on the new {\it AKARI} FSCv1, helped by {\it SDSS}, {\it GALEX}, and {\it WISE} data, demonstrate interesting trends between the E(B$-$V)$_{\rm gas}$/E(B$-$V)$_{\rm star}$ ratio and various galaxy properties. However, considering the potential large uncertainty regarding the derivation of H$\alpha$ SFR (which is inevitable as long as we rely on the SDSS data), more detailed, systematic spatially-resolved studies for a statistical sample of galaxies selected from above/below the SF main sequence would be necessary to obtain a conclusive evidence of the change in the dust geometry across the SFR--$M_{\star}$ diagram.

\begin{ack}

We thank the referee for their helpful and constructive comments which improved the paper. 

This research is based on observations with AKARI, a JAXA project with the participation of ESA. This work was financially supported in part by a Grant-in-Aid for the Scientific Research (Nos.\,26800107; 18K13588) by the Japanese Ministry of Education, Culture, Sports and Science. 

Funding for the SDSS and SDSS-II has been provided by the Alfred P. Sloan Foundation, the Participating Institutions, the National Science Foundation, the U.S. Department of Energy, the National Aeronautics and Space Administration, the Japanese Monbukagakusho, the Max Planck Society, and the Higher Education Funding Council for England. The SDSS Web Site is http://www.sdss.org/. The SDSS is managed by the Astrophysical Research Consortium for the Participating Institutions. The Participating Institutions are the American Museum of Natural History, Astrophysical Institute Potsdam, University of Basel, University of Cambridge, Case Western Reserve University, University of Chicago, Drexel University, Fermilab, the Institute for Advanced Study, the Japan Participation Group, Johns Hopkins University, the Joint Institute for Nuclear Astrophysics, the Kavli Institute for Particle Astrophysics and Cosmology, the Korean Scientist Group, the Chinese Academy of Sciences (LAMOST), Los Alamos National Laboratory, the Max-Planck-Institute for Astronomy (MPIA), the Max-Planck-Institute for Astrophysics (MPA), New Mexico State University, Ohio State University, University of Pittsburgh, University of Portsmouth, Princeton University, the United States Naval Observatory, and the University of Washington.

This research is based on observations made with the NASA Galaxy Evolution Explorer. GALEX is operated for NASA by the California Institute of Technology under NASA contract NAS5-98034.

This publication makes use of data products from the Wide-field Infrared Survey Explorer, which is a joint project of the University of California, Los Angeles, and the Jet Propulsion Laboratory/California Institute of Technology, funded by the National Aeronautics and Space Administration.

\end{ack}




\end{document}